\documentclass[aps, prd, preprint, superscriptaddress, nofootinbib, amsfonts, amssymb, amsmath]{revtex4-1}
\usepackage{ulem, graphicx, bm, color, wrapfig}
\usepackage[colorlinks, linkcolor=red, anchorcolor=blue, citecolor=green]{hyperref}
\usepackage{enumitem}
\begin{document}

\preprint{KEK-TH-2631, KEK-Cosmo-0347}

\title{
% Theoretical and experimental constraints on early-Universe models in \texorpdfstring{$F(R)$}{TEXT} gravity
Theoretical and observational constraints on \\
early dark energy in \texorpdfstring{$F(R)$}{F(R)} gravity
}

\author{Hua Chen}
\email{huachen@mails.ccnu.edu.cn}
\affiliation{
Institute of Astrophysics, Central China Normal University, Wuhan 430079, China
}
\author{Taishi Katsuragawa}
\email{taishi@ccnu.edu.cn}
\affiliation{
Institute of Astrophysics, Central China Normal University, Wuhan 430079, China
}
\author{Shin'ichi Nojiri}
\email{nojiri@nagoya-u.jp}
\affiliation{
KEK Theory Center, Institute of Particle and Nuclear Studies, 
High Energy Accelerator Research Organization (KEK), Oho 1-1, Tsukuba, Ibaraki 305-0801, Japan
}
\affiliation{
Kobayashi-Maskawa Institute for the Origin of Particles and the Universe, 
Nagoya University, Nagoya 464-8602, Japan
}
\author{Taotao Qiu}
\email{qiutt@hust.edu.cn}
\affiliation{School of Physics, Huazhong University of Science and Technology, Wuhan 430074, China}

\begin{abstract}
This work examines an early dark energy (EDE) scenario in the context of $F(R)$ gravity.
EDE is introduced to alleviate the Hubble tension by temporarily injecting approximately $10\%$ of the energy fraction around the matter-radiation equality epoch ($z \approx 10^{3}$--$10^{4}$).
Building on several benchmark models, we focus on the potential-driven EDE scenario and investigate the conditions required within $F(R)$ gravity.
We first introduce a dimensionless quantity to analytically visualize the evolution of the density ratio between EDE and other matter components.
Considering several examples, we demonstrate that the desired EDE can indeed be realized in $F(R)$ gravity.
However, stringent constraints arising from violations of the equivalence principle could exclude the allowed parameter space.
Our result provides a generic constraint on the potential-driven EDE in $F(R)$ gravity at the background level.
This work also concludes that nonperturbative effects or nontrivial mechanisms are indispensable for studying EDE in $F(R)$ gravity while maintaining compatibility with local tests of gravity.
\end{abstract}

\maketitle

%%%%%%%%%%%%%%%%%%%%%%%%%
%%%%%%%%%%%%%%%%%%%%%%%%%
\section{Introduction}

%%%%%%%%%%%%%%%%%%%%%%%%%
% EDE for Hubble tension 
%%%%%%%%%%%%%%%%%%%%%%%%%
The Hubble tension problem, which has been persistent for over a decade, potentially implies new physics in the period of precombination except for the late-time solution~\cite{DiValentino:2021izs, Schoneberg:2021qvd}.
Among various precombination proposals, the early dark energy (EDE) scenario seems a promising one to the Hubble tension~\cite{Kamionkowski:2022pkx}.
This scenario introduces about $10\%$ injection of fractional energy around matter-radiation equality (MRE) and effectively slows down the decay of the Hubble rate, resulting in a decrease of the sound horizon $r_{\mathrm{s}}$~\cite{Kamionkowski:2022pkx}.
Because the angle subtended by the sound horizon $\theta_{\mathrm{s}}=r_{\mathrm{s}}/D_{A}$ is precisely determined by the cosmic microwave background (CMB) measurements, the angular diameter distance decreases accordingly, leading to an increase in the Hubble constant~\cite{Kamionkowski:2022pkx}.

%%%%%%%%%%%%%%%%%%%%%%%%%%%%%%%%%%%%%%%%%
%  EDE as new physics in the dark sector 
%%%%%%%%%%%%%%%%%%%%%%%%%%%%%%%%%%%%%%%%%
The EDE has been examined in a variety of contexts --- for instance,
the dark fluid~\cite{Lin:2019qug, Sabla:2022xzj},
single scalar field~\cite{Braglia:2020bym, Jiang:2021bab},
axionlike scalar field~\cite{Karwal:2016vyq, Poulin:2018dzj},
and two scalar fields~\cite{Niedermann:2019olb}.
In the simplest case of the single scalar field, when the EDE scalar field overcame the Hubble friction at MRE, it started to roll down its potential and decayed faster than radiation.
Such a simple mechanism can significantly alleviate the Hubble tension, though it requires fine-tuning of model parameters.
As a complement, the dark matter (DM)-coupled EDE models provide a natural trigger mechanism: 
When dark matter starts to dominate the Universe at MRE, it activates the dynamics of the EDE scalar field~\cite{Karwal:2021vpk, McDonough:2021pdg, Lin:2022phm}.

%%%%%%%%%%%%%%%%%%%%%%%%%%%%%%%%%%%%
% Modified gravity for Hubble tension
%%%%%%%%%%%%%%%%%%%%%%%%%%%%%%%%%%%%
On the other hand, the modified gravity theory can also be a possible solution to the Hubble tension.
As well as attempts to relieve the Hubble tension by modifying the late-time cosmic evolution~\cite{Benevento:2020fev}, modifications to general relativity (GR) in the early Universe can also influence the CMB measurements~\cite{Lin:2018nxe}.
Since modified gravity theories violate the equivalence principle, they can trigger the time-varying gravitational constant in the early Universe~\cite{Lima:2016npg}.
In light of the Hubble tension, the time-varying gravitational constant during MRE has been investigated in the scalar-tensor theory with a nonminimal coupling~\cite{Ballardini:2020iws, Ballardini:2021evv, Ballesteros:2020sik, Braglia:2020auw} and Horndeski theory~\cite{Zumalacarregui:2020cjh}, 
as well as in the effective field theory approach~\cite{Benevento:2022cql, Kable:2023bsg}.
Moreover, the modified gravity generally introduces additional degrees of freedom to GR, and thus, the EDE scalar field may originate from the modified gravity theory~\cite{SolaPeracaula:2019zsl}.

%%%%%%%%%%%%%%%%%%%%%%%%%%%%%%%%%%%%
% EDE scalar in F(R) gravity
%%%%%%%%%%%%%%%%%%%%%%%%%%%%%%%%%%%%
$F(R)$ gravity theory, one of the modified gravity theories, introduces an additional scalar degree of freedom to GR,
dubbed \textit{scalaron}~\cite{DeFelice:2010aj, Nojiri:2017ncd}.
The scalaron has been extensively studied to explain the early- and late-time cosmic acceleration associated with inflation~\cite{Starobinsky:1980te, DeFelice:2010aj, Codello:2014sua} and late-time dark energy (LDE)~\cite{Hu:2007nk, Starobinsky:2007hu, Appleby:2007vb, Tsujikawa:2007xu, Elizalde:2010ts}.
Moreover, the scalaron couples with the other matter contents in the Universe, and its dynamics are influenced by matter dynamics in a way similar to DM-coupled EDE.  
Existing works have shown the nontrivial scalaron dynamics in the presence of matter~\cite{Erickcek:2013oma, Erickcek:2013dea, Katsuragawa:2017wge, Katsuragawa:2018wbe, Chen:2019kcu, Chen:2022zkc, Shtanov:2021uif, Shtanov:2022xew, Shtanov:2024nmf}.
Given the success of these applications, this work aims to explore the cosmological effects of the scalaron at an intermediate stage, well before and after the two acceleration periods.
In particular, focusing on scalaron as a new scalar field in the dark sector, we study the new physics of precombination in $F(R)$ gravity, for which the scalaron acts as EDE in the same manner as the existing EDE scenarios.

%%%%%%%%%%%%%%%%%%%%%%%%%%%%%%%%%%%%
% Hubble tension in F(R) gravity
%%%%%%%%%%%%%%%%%%%%%%%%%%%%%%%%%%%%
Several works have used $F(R)$ gravity to address the Hubble tension~\cite{Schiavone:2022wvq}.
Nojiri, Odintsov, and Oikonomou~\cite{Nojiri:2019fft} proposed a scenario unifying LDE, EDE, and inflation.
Odintsov \textit{et al.}~\cite{Odintsov:2020qzd, Odintsov:2023cli} utilized the EDE model proposed in Ref.~\cite{Nojiri:2019fft} to study the Hubble tension problem and obtained a Hubble constant similar to the prediction in the $\Lambda$CDM model.
Nojiri, Odintsov, and Oikonomou~\cite{Nojiri:2022ski} proposed a saddle point condition as a remedy for the Hubble tension.
Schiavone, Montani, and Bombacigno~\cite{Schiavone:2022wvq} assumed a special solution of the scalaron field to produce a decreasing trend of the effective Hubble constant based on the idea of redshift-dependent Hubble constant inferred from a binned data analysis of the SNe Ia Pantheon sample within $2\sigma$~\cite{Dainotti:2021pqg, Dainotti:2022bzg}.
They assumed a special solution, the $F(R)$ function at the low-redshift limit, and thereby reconstructed the $F(R)$ function to be $F(R)\approx m^2 B_0+B_1 R+B_2 R^2/m^2$, where $m$ is a constant with the dimension of mass and $B_0$, $B_1$, and $B_2$ are dimensionless constants.
Although the above studies have been phenomenologically successful in light of the Hubble tension, they should also be compatible with other constraints, 
e.g., local tests of gravity~\cite{Capozziello:2007eu}.

%%%%%%%%%%%%%%%%%%%%%%%%%%%%%%%%%%%%
% Brief summary of this work
%%%%%%%%%%%%%%%%%%%%%%%%%%%%%%%%%%%%
Our methodology in this work is to revisit the existing EDE scenarios in terms of the scalaron in $F(R)$ gravity and to investigate the parameter spaces that allow for the injection of a fractional energy density of around $10\%$ around MRE in the range $z=10^{3}$--$10^{4}$.
Under this setup, we clarify all necessary conditions for a viable EDE model in $F(R)$ gravity at the background level.
In particular, we develop a novel theoretical approach by constructing a dimensionless quantity that enables us to visualize the energy-density ratio between the scalaron and other matter contents in a model-independent way.
We then utilize this quantity to study several benchmark models, including existing and new proposed ones.
To investigate the EDE at MRE epoch in a model-independent way, we consider the scenario that the scalaron always follows its effective potential minimum, where the EDE is regarded as the potential energy of the scalaron field.
In this simple scenario, we examine the realization of the EDE in several benchmark models of $F(R)$ gravity, while also discussing the validity of this scenario itself.

%%%%%%%%%%%%%%%%%%%%%%%%%%%%%%%%%%%%
% Brief summary of the result
%%%%%%%%%%%%%%%%%%%%%%%%%%%%%%%%%%%%

We find that these models can indeed produce the required fractional energy injection within suitable regions of the parameter space. 
However, we also show that these regions can be ruled out by local gravity tests. 
Consequently, the parameter space that survives the local gravity constraints severely limits the precombination impact of potential-driven EDE models in $F(R)$ gravity. 
This newly emerging tension indicates a generic constraint on the potential-driven EDE scenario as a resolution to the Hubble tension in $F(R)$ gravity. 
Because the obtained constraint is valid only in the potential-driven EDE scenario, we may also discuss the EDE and early-Universe models of $F(R)$ gravity that, in principle, could evade the proposed constraints.

%%%%%%%%%%%%%%%%%%%%%%%%%%%%%%%%%%%%
% contents
%%%%%%%%%%%%%%%%%%%%%%%%%%%%%%%%%%%%
This paper is organized as follows:
In Sec.~\ref{Background}, we introduce notations by briefly reviewing the basis of $F(R)$ gravity.
In Sec.~\ref{Hubble tension and Early dark energy}, we clarify necessary conditions and construct dimensionless quantities; we then discuss a class of models in each subsection.
In Sec.~\ref{No-go theorem}, we propose generic constraints on the potential-driven EDE and early-Universe models of $F(R)$ gravity.
Finally, we summarize our results and conclude this paper in Sec.~\ref{Conclusion}.

%%%%%%%%%%%%%%%%%%%%%%%%%
%%%%%%%%%%%%%%%%%%%%%%%%%
\section{Background evolution in \texorpdfstring{$F(R)$}{TEXT} gravity}
\label{Background}

The action of $F(R)$ gravity is given by
\begin{align}
    S=\frac{1}{2\kappa^{2}}\int \mathrm{d}^{4}x\,\sqrt{-g}F(R) +S_{\mathrm{Matter}}
    \label{eq: JF action}
    \,,
\end{align}
where $\kappa^{2}\equiv 8\pi G\equiv 1/M_{\mathrm{Pl}}^{2}$, with $M_{\mathrm{Pl}}=2.44\times 10^{18}$ GeV being the reduced Planck mass.
$g$ is the determinant of Jordan-frame metric $g_{\mu\nu}$,
and $S_{\mathrm{Matter}}$ is the matter action describing Standard Model (SM) matter fields and dark matter.
For convenience, we define $F(R)\equiv R+f(R)$ so that $f(R)$ encodes all modifications to the GR.

Variation of the action with respect to the metric yields the modified field equation:
\begin{align}
    G_{\mu\nu}
    = \kappa^{2}T_{\mu\nu} + \kappa^{2}T_{\mu\nu}^{R}
    \label{eq: field eq in Jordan}
    \,,
\end{align}
where $G_{\mu\nu}$ is the Einstein tensor, $G_{\mu\nu} = R_{\mu\nu} - \frac{1}{2} R g_{\mu\nu}$,
$T_{\mu\nu}$ is the energy-momentum tensor for matter fields defined by
\begin{align}
    T_{\mu\nu} 
    \equiv \frac{-2}{\sqrt{-g}} 
    \frac{\delta S_{\mathrm{Matter}}}{\delta g^{\mu\nu}}
    \,,
\end{align}
and $T_{\mu\nu}^{R}$ is the effective energy-momentum tensor for the scalar degree of freedom, dubbed \textit{scalaron} and given by
\begin{align}
    T_{\mu\nu}^{R}
    \equiv
    \frac{1}{\kappa^{2}}\left[\frac{1}{2} \left(F-R\right)g_{\mu\nu}
    +\left(1-F_{R}\right)R_{\mu\nu}
    +\nabla_{\mu}\nabla_{\nu}F_{R}
    -g_{\mu\nu}\Box F_{R}\right]
    \,.
\end{align}
The equation of motion for the scalaron can be found directly from the trace of the field equation \eqref{eq: field eq in Jordan}:
\begin{align}
    3\square F_{R}(R)+F_{R}(R)R-2F(R)=\kappa^{2}T_{~\mu}^{\mu}
    \label{eq: trace eq}
    \,.
\end{align}
Obviously, the higher derivative term $\square F_{R}(R)$ is absent in GR, where $F(R)=R$ and the equation reduces to $R=-\kappa^{2}T_{~\mu}^{\mu}$.

The action~\eqref{eq: JF action} can be rewritten in the form of scalar-tensor theories:
\begin{align}
    S
    =
    \int\mathrm{d}^{4}x\,\sqrt{-g}
    \left[\frac{1}{2\kappa^{2}}\Phi R-U(\Phi)\right]
    + S_{\mathrm{Matter}}
    \,,
\end{align}
where the scalar field $\Phi$,
\begin{align}
    \Phi= F_{R}(R)
    \,,
\end{align}
is directly coupled to the Ricci scalar and the potential is given by
\begin{align}
    U(\Phi)
    =
    \frac{RF_{R}(R)-F(R)}{2\kappa^{2}}
    \,.
\end{align}

Sometimes, it is more intuitive to manifest the scalar degree of freedom in the canonical form:
\begin{align}
\label{eq: einstein action}
    \tilde{S}
    =
    \int d^{4}x\ \sqrt{-\tilde{g}}
    \left[
        \frac{1}{2\kappa^{2}}\tilde{R}
        -\frac{1}{2}\tilde{g}^{\mu\nu}(\tilde{\partial}_{\mu}\varphi)(\tilde{\partial}_{\nu}\varphi)
        -V(\varphi)+e^{-4\sqrt{1/6}\kappa\varphi}\mathcal{L}_{\mathrm{Matter}}
    \right]
    \,,
\end{align}
through the conformal transformation
\begin{align}
\label{eq: Weyl transformation}
    \tilde{g}_{\mu\nu}
    = 
    F_{R}(R)g_{\mu\nu}
    \equiv 
    e^{2\sqrt{1/6}\kappa\varphi}g_{\mu\nu}
    \,.
\end{align}
Here, $\varphi\equiv \sqrt{3/2}M_\mathrm{Pl}\ln F_R$ is a new scalar field defined in the Einstein frame, and the potential is
\begin{align}
\label{eq: scalaron potential}
    V(\varphi)
    =
    \frac{1}{2\kappa^{2}}\frac{RF_R(R)-F(R)}{F_{R}^{2}(R)}
    =
    \frac{U(\Phi)}{F_{R}^{2}(R)}
    \,.
\end{align}

The energy-momentum tensor in both frames is related by  $\tilde{T}_{\,\nu}^{\mu} = F_{R}^{-2}T_{\ \nu}^{\mu}$, and so are the energy density $\tilde{\rho}=F_{R}^{-2}\rho$ and pressure $\tilde{P}=F_{R}^{-2}P$. 
We note that the equation-of-state parameter $w=P/\rho$ is frame independent.
Assuming $w$ is constant, the energy density conserved in the Einstein frame is given by $\tilde{\rho}_{\mathrm{c}}\equiv\rho F_R^{-3(1+w)/2}$ rather than $\tilde{\rho}$.
Note that the two frames are equivalent, but we should be careful about the physical interpretations.
However, because we are concerned about only the cosmic history after big bang nucleosynthesis (BBN), our analysis is almost frame independent, as will be discussed in Sec.~\ref{subsec: Power-law EDE}. 

The field equation in the Einstein frame is
\begin{align}
    \tilde{R}_{\mu\nu} - \frac{1}{2}\tilde{R}\tilde{g}_{\mu\nu}
    &=
    \kappa^{2}\tilde{T}_{\mu\nu}^{(\phi)}
    +\kappa^{2}\tilde{T}_{\mu\nu}
    \,,
\end{align}
where
\begin{align}
    \tilde{T}_{\mu\nu}^{(\phi)}
    \equiv 
    -\frac{1}{2} \tilde{g}_{\mu\nu}\tilde{g}^{\rho\sigma}
    (\tilde{\partial}_{\rho}\varphi)(\tilde{\partial}_{\sigma}\varphi)
    +(\tilde{\partial}_{\mu}\varphi)(\tilde{\partial}_{\nu}\varphi)
    -\tilde{g}_{\mu\nu}V(\varphi)
    \,.
\end{align}
In the flat Friedmann-Lema\^{i}tre-Robertson-Walker spacetime, the energy density and pressure of the scalaron are then 
\begin{align}
    \tilde{\rho}_\varphi
    &=
    \frac{1}{2} \left(\frac{d\varphi}{d\tilde{t}}\right)^{2}
    + V(\varphi)
    \,,\\
    \tilde{P}_\varphi
    &=
    \frac{1}{2}\left(\frac{d\varphi}{d\tilde{t}}\right)^{2}
    -V(\varphi)
    \,,
\end{align}
respectively, where $\tilde{t}$ is the cosmic time in the Einstein frame.
The Friedmann equation reads
\begin{align}
    3M_{\mathrm{Pl}}^{2}\tilde{H}^{2}
    &=
    \tilde{\rho}_{\varphi} + \tilde{\rho}_{\mathrm{m}}+\tilde{\rho}_{\mathrm{r}}
    \,,
\end{align}
where $\tilde{\rho}_\mathrm{m}$ and $\tilde{\rho}_\mathrm{r}$ are the energy densities for matter (baryons and dark matter) and radiation, respectively.
Moreover, using the pressures for the matter and radiation, $\tilde{P}_\mathrm{m}$ and $\tilde{P}_\mathrm{r}$, we define the total energy density and pressure as, respectively, 
\begin{align}
    \tilde{\rho}_{\mathrm{tot}} 
    = 
    \tilde{\rho}_{\varphi} + \tilde{\rho}_{\mathrm{m}}+\tilde{\rho}_{\mathrm{r}}
    \, , \ 
    \tilde{P}_{\mathrm{tot}} 
    = 
    \tilde{P}_{\varphi} +\tilde{P}_{\mathrm{m}}+\tilde{P}_{\mathrm{r}}
    \, .
\end{align}
It can be conveniently rewritten as
\begin{align}
    \Omega_{\varphi}+\Omega_{\mathrm{m}}+\Omega_{\mathrm{r}}=1
    \,,
\end{align}
where we have defined the density parameter
\begin{align}
    \Omega_{i}
    \equiv
    \frac{\tilde{\rho}_{i}}{\tilde{\rho}_\mathrm{cr}}
    \equiv
    \frac{\tilde{\rho}_{i}}{3M_{\mathrm{Pl}}^{2}\tilde{H}^{2}}\,.
\end{align}

In the Einstein frame, the equations of motion for matter fields and the scalaron are a coupled system:
\begin{align}
    \frac{d\tilde{\rho}}{d\tilde{t}}
    +3\tilde{H}(\tilde{\rho}+\tilde{P})
    &=
    -\frac{\kappa}{\sqrt{6}}(\tilde{\rho}
    -3\tilde{P})\frac{d\varphi}{d\tilde{t}}
    \,,\\
    \frac{d^{2}\varphi}{d\tilde{t}^{2}}
    +3\tilde{H}\frac{d\varphi}{d\tilde{t}}
    +V_{,\varphi}
    &=
    \frac{\kappa}{\sqrt{6}}(\tilde{\rho}-3\tilde{P})\,,
\end{align}
where $\tilde{\rho}$ and $\tilde{P}$ are the total energy density and pressure of the matter contents, respectively. 
The slope of the potential is given as
\begin{align}
    V_{,\varphi}(\varphi)
    =
    \frac{1}{\sqrt{6}\kappa}\frac{2F(R)-RF_{R}(R)}{F_{R}^{2}(R)}
    \,.
\end{align}
Using the two equations of motion, we find that the total energy density is conserved:
\begin{align}
\frac{d\tilde{\rho}_{\mathrm{tot}}}{d\tilde{t}}+3\tilde{H}\left(\tilde{\rho}_{\mathrm{tot}}+\tilde{P}_{\mathrm{tot}}\right)=0
\,.
\end{align}
Finally, the effective mass squared of the scalaron is defined by
\begin{align}
\label{m2}
\begin{split}
    m_{\varphi}^{2}
    &\equiv 
    V_{\mathrm{eff},\varphi\varphi}(\varphi_{\min})
    \\
    &=
    \frac{1}{3F_{RR}(R_{\min})}
    \left[1-\frac{R_{\min}F_{RR}(R_{\min})}{F_{R}(R_{\min})}\right]
    \,,
\end{split}
\end{align}
where the effective potential $V_{\mathrm{eff}}$ includes the matter contribution:
\begin{align}
    V_{\mathrm{eff}, \varphi}(\varphi)
    =
    \frac{1}{\sqrt{6}\kappa}
    \left[ \frac{2F(R)-RF_{R}(R)}{F_{R}^{2}(R)}
    - \kappa^2(\tilde{\rho}-3\tilde{P})
    \right]
    \,.
\end{align}
$R_\mathrm{min}$ is the Ricci scalar corresponding to the effective potential minimum,
which evolves in time as $T_{~\mu}^{\mu}$ does.

%%%%%%%%%%%%%%%%%%%%%%%%%
%%%%%%%%%%%%%%%%%%%%%%%%%
\section{Hubble tension and Early dark energy}
\label{Hubble tension and Early dark energy}

%%%%%%%%%%%%%%%%%%%%%%%%%
\subsection{Potential-driven EDE tracking matter evolution}

Before constructing the EDE scenario in $F(R)$ gravity, we first consider the scalaron dynamics in the early Universe.
Here, the early Universe refers to the period much earlier than the LDE-dominant epoch, denoted by $R\gg R_\mathrm{LDE}$ in terms of the curvature.
The equation of motion for the scalaron~\eqref{eq: trace eq} includes
the trace of the energy-momentum tensor of matter, $T_{~\mu}^{\mu}=-\rho+3P$, which behaves as an effective external force term.
Unlike the EDE scalar field in the pure dark sector or the DM-coupled EDE field in Ref.~\cite{Karwal:2021vpk}, the scalaron is universally coupled to matter, including baryons and dark matter, in $F(R)$ gravity.
Therefore, the scalaron's dynamics differ from those of other EDE scenarios.
As shown in Ref.~\cite{Chen:2022zkc}, the trace term did not vanish even though $-T_{~\mu}^{\mu}\ll \rho_\mathrm{r}$ deep into the radiation-dominated epoch.
Momentarily, the successive decoupling effect of the SM particles from the thermal bath would make the scalaron overcome the Hubble friction and drive it to its effective potential minimum, even though the scalaron overshot or was misaligned somewhere far away from its equilibrium value.

At this point, we assume that the background evolution of the scalaron can be traced through that of the effective potential minimum.
In other words, we study only the spatially averaged field minimum $\varphi_\mathrm{min}=\varphi(R_\mathrm{min})$ evaluated by $R_{\mathrm{min}}$ through the stationary condition $V_{\mathrm{eff}, \varphi}(\varphi) = 0$:
\begin{align}
\label{eq: stability condition}
    R_{\mathrm{min}}F_{R}\left(R_{\mathrm{min}}\right)
    -2F\left(R_{\mathrm{min}}\right)
    =\kappa^{2}T_{~\mu}^{\mu}
    \,.
\end{align}
Here, we impose $F(R)>0$ for $R>R_{0}$, where $R_{0}$ is a curvature scale at the lower bound of the adaptive range of the model.
It is natural that $R_{0}$ should be associated with the LDE scale if we consider the whole cosmic history.
We also require $F_R>0$ and $F_{RR}>0$ to avoid ghost and tachyonic instabilities for $R>R_{0}$.
From a theoretical perspective, we shall constrain model parameters using these three \textit{stability conditions}, independently of phenomenological requirements for the Hubble tension.

Regarding the Hubble tension problem, we expect a non-negligible contribution of the scalaron energy to the cosmic-energy budget between $z=10^{3}$ and $z=10^{4}$, which peaks at MRE.
Using $\rho_\varphi = 0.1\tilde{\rho}_\mathrm{tot}$ as a reference value for the scalaron energy density, we impose each component at MRE to satisfy the following ratio:
\begin{align}
    \tilde{\rho}_{\varphi}: \tilde{\rho}_{\mathrm{m}}: \tilde{\rho}_{\mathrm{r}}
    \approx 
    0.1: 0.45: 0.45
    \,.
\end{align}
Here, the scalaron energy density $\tilde{\rho}_{\varphi}$ is assumed to account for about $10\%$ of the total energy density as EDE.
The remaining $90\%$ is shared equally between the matter and radiation energy densities.

Hereafter, we focus on the potential-driven EDE scenarios in $F(R)$ gravity.
Thus, we ignore the kinetic term of the scalaron, and the dynamics is characterized by the potential $V(\varphi)$ or, equivalently, $F(R)$ function.
Note that, in the sense of ignoring the kinetic term, our scenario seems analogous to the conventional slow-roll inflation and LDE.
However, the scalaron does not roll on the potential; instead, {\it the shape of the effective potential evolves as the scalaron stays at the potential minimum}.
The scalaron dynamics is, thus, determined by the stationary condition \eqref{eq: stability condition}, following the dynamics of the other matter contents. 
Although we investigate the background cosmic evolution in this specific scenario, the formulation applies to general models of $F(R)$ gravity. 
In this scenario, if the scalaron plays the role of EDE, its potential energy dominates over its kinetic energy, leading to $\rho_\varphi\approx V(\varphi)$.
The density ratio of the scalaron and matter at MRE reads
\begin{align}
\label{eq: energy ratio}
    \frac{V(\varphi_{\mathrm{min}})}{\tilde{\rho}_{\mathrm{m}}} 
    \approx \frac{2}{9}
    \approx 0.222
    \,.
\end{align}
Note that if we allow $\pm5\%$ deviations in the scalaron energy density $\tilde{\rho}_{\varphi}$, Eq.~\eqref{eq: energy ratio} leads to $2/19~(\approx 0.105) \lesssim V(\varphi_{\mathrm{min}})/\tilde{\rho}_{\mathrm{m}} \lesssim 6/17~(\approx 0.353)$.

On the other hand, combining Eqs.~\eqref{eq: scalaron potential} and~\eqref{eq: stability condition} yields the density ratio:
\begin{align}
\label{eq: density ratio}
\begin{split}
    \frac{V(\varphi_{\mathrm{min}})}{-\tilde{T}_{~\mu}^{\mu}}
    &=
    \frac{U(\Phi_{\mathrm{min}})}{-T_{~\mu}^{\mu}}
    \approx
    \frac{U(\Phi_{\mathrm{min}})}{\rho_{\mathrm{m}}}
    =
    \frac{V(\varphi_{\mathrm{min}})}{\tilde{\rho}_{\mathrm{m}}}
    \\
    &=
    \frac{1}{2}
    \left[
        \frac{1}{2-R_{\mathrm{min}}F_{R}(R_{\mathrm{min}})/F(R_{\mathrm{min}})}-1
    \right]
    \,.
\end{split}
\end{align}
The first equality reflects that dimensionless quantities are frame independent, as the equation-of-state parameter $w$, so our discussion applies to both frames.
The approximation $-T_{~\mu}^{\mu}\approx\rho_\mathrm{m}$ is valid after $e^{+}e^{-}$ annihilation, occurring much earlier than MRE. 
Hence, it applies to our current consideration.
The second line implies that the dimensionless density ratio can be translated into another dimensionless quantity defined by
\begin{align}
\label{eq: r}
\begin{split}
    r(R_{\mathrm{min}})
    &\equiv 
    \frac{R_{\mathrm{min}}F_{R}(R_{\mathrm{min}})}{F(R_{\mathrm{min}})}
    \\
    &\approx 
    1+\frac{2V(R_{\mathrm{min}})/\tilde{\rho}_{\mathrm{m}}}{2V(R_{\mathrm{min}})/\tilde{\rho}_{\mathrm{m}}+1}
    \,.
\end{split}
\end{align}
It is more convenient to use $r$ rather than the density ratio, because $r$ is a bounded quantity, i.e., $1\leq r\leq 2$.
We obtain $r\approx1$ when the scalaron potential energy and modification to GR are negligible, and $r\approx2$ when they are dominant, specifically, $r\approx 1$ at the matter-dominated epoch 
[$V(\varphi_{\mathrm{min}}) \ll \tilde{\rho}_{\mathrm{m}}$]
and $r\approx2$ near the de Sitter point
[$V(\varphi_{\mathrm{min}}) \gg \tilde{\rho}_{\mathrm{m}}$].
In order for $V/\tilde{\rho}_\mathrm{m}\geq 2/9$ at MRE, one requires $r\geq 17/13$~\footnote{
If we apply it to the LDE model in $F(R)$ gravity, 
we find $r\approx 31/17$ at present Universe.
[$V(\varphi_{\mathrm{min}}) : \tilde{\rho}_{\mathrm{m}} \approx 7 : 3$]
}.
Note that for the $5\%$ deviation in $V/\rho_m$, $r$ is in the range of $27/23 \leq r\leq 41/29$.

\subsection{
Validity of potential-driven scenario and adiabatic condition
}

We have assumed that the background evolution of the scalaron follows the evolution of the stationary condition, which can be rephrased as the scalaron staying at the minimum of the effective potential.
As a result, the kinetic energy of the scalaron is ignored, but the scalaron field evolves in time as the potential structure and its minimum evolve.
Thus, in the potential-driven EDE scenario, we have conditions on the time derivative of the scalaron field and its potential.
In this subsection, we discuss the validity of the potential-driven EDE scenario in terms of the perturbation.

We now consider the scalaron dynamics in the presence of a time-dependent potential minimum $\varphi_0(t)$.
If the scalaron is slightly displaced from the evolving minimum, its evolution is governed by the deviation from the minimum and by the time dependence of the potential structure.
We introduce a linear perturbation $\delta \varphi(t) = \varphi(t) - \varphi_0(t)$, whose equation of motion is given by
\begin{align}
    \delta \ddot{\varphi} + 3 H \delta \dot{\varphi} + m_{\mathrm{eff}}^2(t) \delta \varphi
    = 
    - \ddot{\varphi}_0 - 3 H \dot{\varphi}_0 
    \,.
\end{align}
The right-hand side represents a forcing term induced by the time-dependent potential minimum.

The scalaron remains close to the instantaneous minimum if the relative displacement is sufficiently small.
Thus, in the regime where the timescale of the scalaron oscillation is much shorter than that of the minimum evolution, the forced-oscillator equation admits a quasistatic approximate solution, from which the following estimate is obtained:
\begin{align}
    \left| \frac{\delta \varphi}{\varphi_0} \right| 
    \approx
    \frac{|\ddot{\varphi}_0 + 3 H \dot{\varphi}_0|}{m_{\mathrm{eff}}^2 |\varphi_0|} \ll 1 \,.
\end{align}
This inequality constitutes the adiabatic condition under which the scalaron follows the evolving minimum.
Equivalently, when the effective mass $m_{\mathrm{eff}}$ is sufficiently large compared to the timescale over which the potential evolves, the system remains close to the instantaneous equilibrium point.
Note that higher-order time derivatives of $\varphi_0(t)$ may be included for more refined estimates, but they enter only at higher order in the adiabatic expansion and do not alter the qualitative conclusion.

Therefore, within the potential-driven scenario adopted in this work, the scalaron can adiabatically track the evolving minimum of the effective potential.
In this sense, the adiabatic condition is imposed as a working assumption to provide a model-independent analysis of the potential-driven EDE scenario.
Nevertheless, as will be discussed later, independent observational considerations require a large effective mass, which naturally ensures a clear separation of timescales and validates the adiabatic approximation assumed here.

%%%%%%%%%%%%%%%%%%%%%%%%%
\subsection{Constraints on $F(R)$ gravity}

Next, we examine the observational constraints on the potential-driven EDE scenario in the $F(R)$ gravity.
Given the constraints, we need to carefully consider the consistency with our current assumption of the stationary condition.
When we assume that the scalaron field stays at the minimum of the time-evolving potential, the scalaron evolution can be treated as quasistatic.
That is, the observational constraints derived under the assumption of a static background spacetime in the late-time Universe can be applied to our potential-driven EDE scenario at a given epoch in the early Universe.
For instance, we consider the fifth-force constraints around the test bodies, which are modeled on the static background, and these constraints at specific curvature scales in the current Universe can be converted into those at certain epochs in the early Universe.
In other words, it is unnecessary to rely solely on constraints from the early-Universe observations, even though our focus is on the EDE scenario at the MRE epoch.
Note that the curvature scale is also converted into the matter density scale through the stationary condition.
By collecting the constraints in the order of the matter density, instead of the curvature, one obtains the constraints on the $F(R)$ function at the static backgrounds.

Robust and stringent constraints on $F_R$ come from the violation of the equivalence principle, and these observations and experiments have been evaluated in the current Universe.
The scalaron-mediated fifth force around the spherically symmetric body constrains the model parameters in the $F(R)$ gravity.
Theoretical analysis of the fifth force primarily relies on the perturbative approach; that is, the field equation for the scalaron is linearized around the potential minimum determined by the stationary condition.
Thus, we can use these constraints in our analysis, correlating the matter density input in the fifth-force experiments and observations with that at a given epoch in cosmic history. 
Moreover, the large effective mass of the scalaron plays a vital role in considering the fifth force. 
If the scalaron field is heavy enough, the scalaron-mediated fifth force is naturally screened, which is called the chameleon mechanism~\cite{Brax:2004qh}.
As was discussed in the previous subsection, the large effective mass is indeed consistent with the potential-driven EDE scenario.

The assumption that the scalaron follows the effective potential minimum allows us to estimate the field amplitude $F_R(R_{\mathrm{min}}) = e^{2\sqrt{1/6}\kappa\varphi_{\min}}$ by comparing energy densities in different environments.
Roughly, the fifth-force experiment suggests $F_R\simeq 1$ for the matter density ranging from atmosphere $\rho_{\mathrm{atm}}\simeq 10^{-3}\,\mathrm{g\cdot cm^{-3}}$ to galaxy $\rho_\mathrm{galaxy}\simeq 10^{-24}\,\mathrm{g\cdot cm^{-3}}$~\cite{Capozziello:2007eu}.
For the energy density at the MRE, $\rho_{\mathrm{eq}}$, since $\rho_\mathrm{galaxy}\ll \rho_\mathrm{eq}\ll \rho_\mathrm{atm}$, we must have $F_R(R_\mathrm{eq})\simeq 1$ provided that $F_R(R)$ is an increasing function of $R$ induced by the stability conditions.

Apart from local gravity tests, we can confront the EDE model with the BBN constraint on the scalaron-field amplitude.
A mass scale $m_i$ would get modified by the scalaron as $\tilde{m}_i=m_i e^{-\sqrt{1/6}\kappa\varphi}$.
To respect the successful BBN based on the SM prediction, a conservative requirement on the variation of the mass scale from the time of BBN to today should be less than $10\%$~\cite{Brax:2004qh}; that is,
\begin{align}
    \left|\frac{m_i\left(\varphi_{\mathrm{BBN}}\right)-m_i\left(\varphi_{\mathrm{today}}\right)}{m_i\left(\varphi_{\mathrm{today}}\right)}\right|\lesssim10\%
    \,,
\end{align}
which implies that
\begin{align}
    \left(\frac{10}{11}\right)^{2}\lesssim F_{R}(R_{\mathrm{BBN}})\lesssim\left(\frac{10}{9}\right)^{2}
    \label{eq: BBN bound}
    \,.
\end{align}
We will revisit the above arguments about our considerations in applying the experimental and observational constraints in Sec.~\ref{No-go theorem}.

%%%%%%%%%%%%%%%%%%%%%%%%%
%%%%%%%%%%%%%%%%%%%%%%%%%
\section{EDE models in \texorpdfstring{$F(R)$}{TEXT} gravity and constraints}

%%%%%%%%%%%%%%%%%%%%%%%%%
\subsection{Power-law EDE}
\label{subsec: Power-law EDE}

We have constructed a dimensionless quantity $r(R_{\mathrm{min}})$ to quantify the density ratio, assuming the scalaron follows the potential minimum in the period of interest.
In the following subsections, we shall apply this quantity to study several classes of EDE models.
We first consider the model with a power-law form given by
\begin{align}
    F(R) 
    &= 
    R + f_{\mathrm{ede}}(R)
    \, , \\
\label{eq: power-law model}
    f_\mathrm{ede}(R) 
    &= 
    \frac{m_{0}^{2}}{3n\left(n-1\right)} \left( \frac{R}{m_{0}^{2}} \right)^{n}
    \,,
\end{align}
where $m_{0}$ is a mass scale to be determined.
This model describes the inflation~\cite{DeFelice:2010aj, Codello:2014sua}
and accommodates Starobinsky's inflation model with $n=2$~\cite{Starobinsky:1980te} if the mass scale is promoted to the inflation scale.
$R^n$ correction also appears in the weak field limit assuming asymptotically Schwarzschild spacetime~\cite{Numajiri:2021nsc, Scali:2024ftw}.

\begin{figure}[htbp]
    \centering
    \includegraphics[scale=0.8]{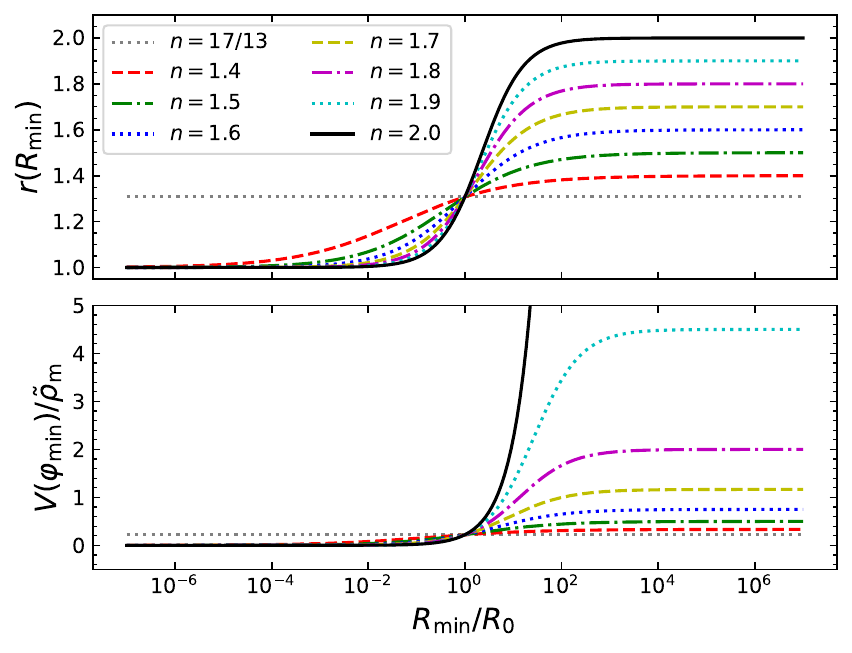}
    \caption{
    The upper panel shows $r$ versus $R_\mathrm{min}/R_{0}$, where we have fixed $R_{0}$ by choosing the equality in Eq.~\eqref{eq: M2R0 power-law}.
    We find $r\to 1$ for $R\ll R_{0}$, implying a vanishing $R^n$ term, and $r\to n$ for $R\gg R_{0}$, implying a tracker solution of the scalaron regarding matter density.
    The lower panel shows $\tilde{\rho}_\mathrm{m}$, with the same parameter choices as the upper panel.
    It tends to vanish for $R\ll R_{0}$, satisfying the requirement of the matter-dominated epoch, and it approaches a fixed value for $R\gg R_{0}$ [see Eq.~\eqref{eq: density_ratio_Rn large R}], indicating a tracker solution.}
    \label{fig: r_rho_Rn}
\end{figure}

\begin{figure}[htbp]
    \centering
    \includegraphics[scale=0.7]{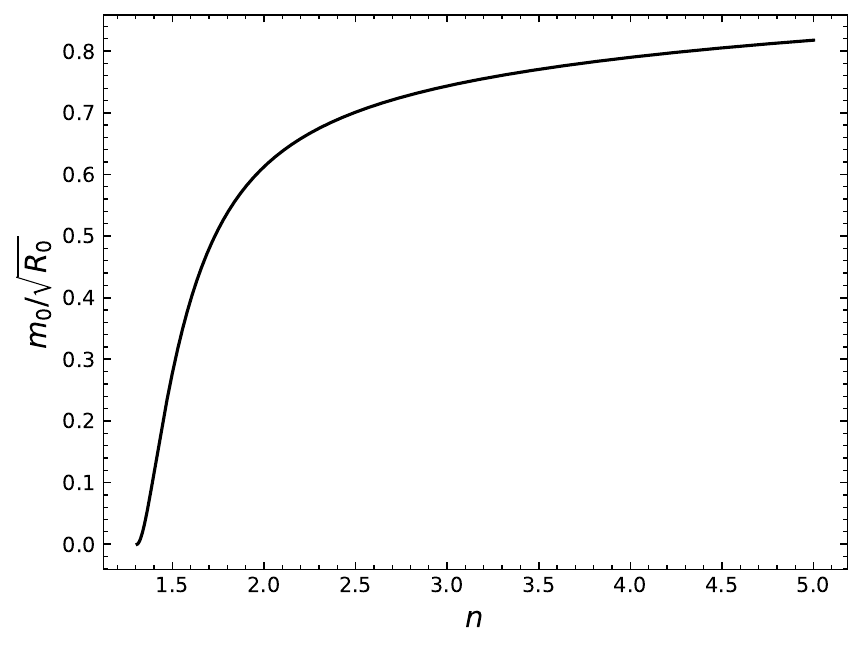}
    \caption{
    $m_{0}$ normalized by $\sqrt{R_0}$ as a function of $n$, where we have used the equality in Eq.~\eqref{eq: M2R0 power-law}.
    $m_{0}/\sqrt{R_0}\to 1$ as $n\to\infty$, and $m_{0}/\sqrt{R_0}\to 0$ as $n\to 17/13$.
    It shows that $m_{0}$ must be less than $\sqrt{R_0}$ to satisfy the condition~\eqref{eq: M2R0 power-law}.}
    \label{fig: m0_R0_n}
\end{figure}

The stability condition, $F_{R}>0$ for $R>0$, constrains the parameter as $n>1$.
Substituting Eq.~\eqref{eq: power-law model} into Eq.~\eqref{eq: r} gives
\begin{align}
    r&=\frac{3n\left(n-1\right)+n\left(R_{\mathrm{min}}/m_{0}^{2}\right)^{n-1}}{3n\left(n-1\right)+\left(R_{\mathrm{min}}/m_{0}^{2}\right)^{n-1}}\,.
    \label{eq: r-Rn}
\end{align}
The asymptotic behavior of $r$ reads
\begin{align}
r(R_{\mathrm{min}})
&=\begin{cases}
1+\frac{1}{3n}\left(R_{\mathrm{min}}/m_{0}^{2}\right)^{n-1}\,, & R_{\mathrm{min}}\ll m_{0}^{2}\,.\\
n+3n\left(n-1\right)\left(1-n^{n-1}\right)\left(m_{0}^{2}/R_{\mathrm{min}}\right)^{n-1}\,, & R_{\mathrm{min}}\gg m_{0}^{2}\,.
\end{cases}
\end{align}
It shows that the scalaron becomes negligible ($r\to 1$) for $R_{\mathrm{min}}\ll m_{0}^{2}$,
while $R_{\mathrm{min}}\gg m_{0}^{2}$, $r\to n$, implying a fixed density ratio at large curvature:
\begin{align}
    \left.\frac{V}{\tilde{\rho}_{\mathrm{m}}}\right|_{R_{\mathrm{min}}\gg m_{0}^{2}}\approx\frac{n-1}{2\left(2-n\right)}\,.
    \label{eq: density_ratio_Rn large R}
\end{align}
To satisfy the condition $r(R_{\mathrm{min}}=R_{0})\geq 17/13$, the mass scale is constrained to be
\begin{align}
    \frac{m_{0}^{2}}{R_{0}}\leq\left[\frac{13n-17}{12n\left(n-1\right)}\right]^{\frac{1}{n-1}}
    \,,
    \label{eq: M2R0 power-law}
\end{align}
where $R_0$ is the critical Ricci scalar corresponding to $V/\tilde{\rho}_\mathrm{m}=2/9$.
As shown in Fig.~\ref{fig: r_rho_Rn}, the EDE decays quickly when $R < R_0$ and tracks the matter evolution when $R > R_{0}$.
However, the condition~\eqref{eq: M2R0 power-law} constrains the mass scale to be small.
As shown in Fig.~\ref{fig: m0_R0_n}, it is a monotonically increasing function and approaches one as $n$ goes to infinity, suggesting that $m_{0}^2\leq R_0$ for all $n\geq 17/13$.


\begin{figure}[htbp]
    \centering
    \includegraphics[scale=0.7]{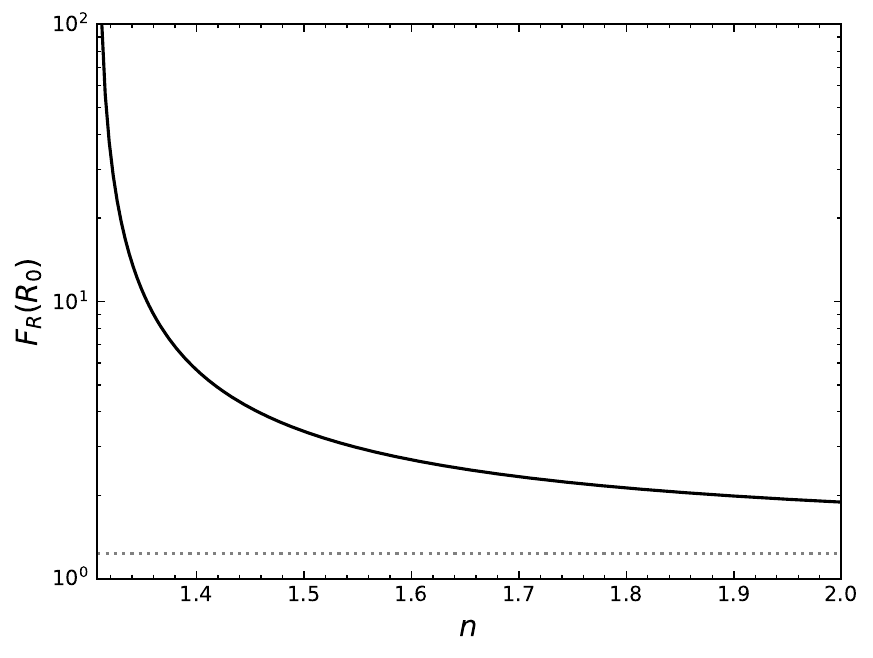}
    \caption{
        The solid black line shows $F_R(R_0)$ as a function of $n$ as in Eq.~\eqref{eq: FR(R0)_n}, and the dotted line represents $F_R(R_0)=(10/9)^2$, which is the upper bound corresponding to BBN. One observes that $F_R(R_0)$ diverges at $n=17/13$ and $F_R(R_0) \to 17/13$ as $n\to\infty$. 
        Therefore, it never goes below the dotted line for $n>17/13$.
        }
    \label{fig: FR(R0)_n}
\end{figure}

If we fix $m_{0}$ by translating the experimental upper bound on the interaction range of the fifth force into the lower bound of the scalaron mass, 
$m_{0}\gtrsim 3\times10^{-3}$~eV~\cite{Lee:2020zjt},
$R_0$ should satisfy the condition~\eqref{eq: M2R0 power-law}, and the required value is too large to be relevant to the precombination phenomenon.
On the other hand, to relieve the Hubble tension, we require $R_{0}$ to be in the same order of Ricci scalar at MRE, i.e., $R_{0}\approx R_\mathrm{eq}$.
Then, $m_{0}\leq\sqrt{R_\mathrm{eq}}$ according to the condition~\eqref{eq: M2R0 power-law}.
One observes that the first derivative of the $F(R)$ function,
\begin{equation}
\begin{aligned}
    F_{R}
    &=1+\frac{1}{3\left(n-1\right)}\left(\frac{R}{m_{0}^{2}}\right)^{n-1}
    \\
    &\geq 1+\frac{4n}{13n-17}\left(\frac{R}{R_\mathrm{eq}}\right)^{n-1}
    \,,
    \label{eq: F_R for R^n}
\end{aligned}
\end{equation}
will be much larger than unity when $R_{\mathrm{min}}>R_\mathrm{eq}$.
Applying the BBN constraint~\eqref{eq: BBN bound}, since $1<F_R(R_\mathrm{eq})<F_R(R_\mathrm{BBN})$, we obtain
\begin{align}
    1
    <F_R(R_0)=\frac{17\left(n-1\right)}{13n-17}
    <\left(\frac{10}{9}\right)^{2} \approx 1.23
    \label{eq: FR(R0)_n}
    \,.
\end{align}
As shown in Fig.~\ref{fig: FR(R0)_n}, this condition cannot be satisfied for any choices of parameter $n>17/13\approx 1.31$.
It should be noted that the above BBN constraint is model independent since it stems from the frame transformation of $F(R)$ gravity theory.

Finally, for the complete analysis, we investigate another power-law model with the opposite sign for Eq.~\eqref{eq: power-law model}:
\begin{align}
\label{eq: power-law model2}
    f_\mathrm{ede}(R) = -c m_{0}^2\left(\frac{R}{m_{0}^2}\right)^{n}
    \,,
\end{align}
where $c>0$.
Note that Eq.~\eqref{eq: power-law model2} represents a DE model if the mass scale $m_{0}^2$ is of the order of the LDE vacuum curvature.
And it is constrained that $n\ll 1$ for this model to be viable as DE~\cite{Capozziello:2007eu}.
Demanding the stability conditions $F, F_R, F_{RR}>0$ for $R>R_{1}$, from theoretical perspectives, 
we find that these constrains lead to $R>R_{1}\gtrsim c^{\frac{1}{1-n}}m_{0}^{2}$, $R>R_{1}\gtrsim\left(cn\right)^{\frac{1}{1-n}}m_{0}^{2}$, and $0<n<1$, respectively.
In order for this model to be valid as an EDE model, $R_\mathrm{eq}\gg R_{1}$, which leads to
\begin{align}
    r(R_{\mathrm{min}})
    &=
    \frac{1-cn\left(\frac{m_{0}^{2}}{R_{\mathrm{min}}}\right)^{1-n}}{1-c\left(\frac{m_{0}^{2}}{R_{\mathrm{min}}}\right)^{1-n}}
    \simeq 
    1
\end{align}
for $R_{\mathrm{min}}\sim R_\mathrm{eq}$.
This result is inconsistent with $r\geq 17/13$ for the EDE, and, thus, the power-law model with a negative sign is also not a consistent EDE model.

%%%%%%%%%%%%%%%%%%%%%%%%%
\subsection{Saddle EDE}

In the previous section, we demonstrated that the power-law term could not apply to the MRE epoch since its mass scale must be large ($m_{0}\gg\sqrt{R_\mathrm{eq}}$) in order to suppress the growth of $F_{R}$.
Reference~\cite{Nojiri:2022ski} suggests that the Hubble tension requires a saddle in the potential $V(R)$.
Hence, this section aims to develop possible saddle EDE models, i.e., broken power-law models.
We propose a general EDE model with a saddle potential $V(R(\varphi))$, written as
\begin{align}
    f_{\mathrm{ede}}(R)
    =\frac{R^{n}\left(c_{1}R-c_{2}R_{0}\right)}{c_{3}R^{n}+R_{0}^{n}}
    \,,
    \label{general EDE}
\end{align}
where $c_1, c_2>0$ are dimensionless coefficients and we fix the parameter as $R_{0}=4\times10^{10}\Lambda$ to roughly correspond to the Ricci scalar at MRE ($R_\mathrm{eq}$), with $\Lambda$ being the cosmological constant; $c_{3}\sim\mathcal{O}(1)$ is, therefore, introduced to slightly shift $R_{0}$ to match $R_\mathrm{eq}$.

This model is similar to the Hu-Sawicki DE model; the difference is that $R$ should evolve from $R\gg R_0$ to $R\ll R_0$, while in the Hu-Sawicki model $R>R_\mathrm{c}$ for model stability, where $R_\mathrm{c}$ is the critical curvature of the same order as the current background curvature.
Therefore, higher powers of the Ricci scalar in EDE are highly restricted, since it may lead to $f_{RR}(R)<0$ at $R\sim R_{0}$.
In this sense, a proper choice is $n=1$, and one can find two base models classified from their signs:
\begin{align}
\text{Negative EDE:}\quad
    f_{-}(R)=-R_{0}\frac{c_{2}R}{c_{3}R+R_{0}}
    \,.
    \label{eq: ede mimus}
\end{align}
and
\begin{align}
\text{Positive EDE:}\quad
    f_{+}(R)=\frac{c_{1}R^{2}}{c_{3}R+R_{0}}
    \,,
    \label{eq: ede plus}
\end{align}
Generally speaking, there are many possible saddle models except for model~\eqref{general EDE}; however, the basic properties are the same and are contained in the above two cases.
We now investigate them separately.

%%%%%%%%%%%%%%%%%%%%%%%%%
\subsubsection{Case 1: \texorpdfstring{$f_{\mathrm{ede}}<0$}{TEXT}}

We first study the negative EDE~\eqref{eq: ede mimus} that is also known as the Hu-Sawicki or Starobinsky model of lowest power with the LDE curvature scale characterized by $R_{0}$.
Likewise, it behaves as an effective cosmological constant for $R\gg R_{0}$.
This model was first discussed in Ref.~\cite{Nojiri:2019fft} and has been utilized to resolve the Hubble tension~\cite{Odintsov:2020qzd, Odintsov:2023cli}.
However, it fails to alleviate the tension compared to the $\Lambda$CDM model.
The model used in Ref.~\cite{Odintsov:2023cli} is given as
\begin{align}
    f_{\mathrm{ede}}(R) = -2\Lambda\alpha\frac{R}{R+R_{0}}
    \,,
\end{align}
where $R_{0}/2\Lambda\equiv 10^{8}$ responsible for the recombination epoch.
Note that $\Lambda$ does not denote the cosmological constant for the late-time Universe.
Reference~\cite{Odintsov:2023cli} concluded $\log \alpha=7.64^{+4.45}_{-2.62}$ with 1$\sigma$ confidence level inferred from the Markov chain Monte Carlo sampling.
Correlating the parameters $c_2 R_0 = 2\Lambda \alpha$ in our notation, we obtain $c_2 = (R_0/2\Lambda)^{-1} \cdot \alpha$.
Thus, we obtain $c_2=10^{-0.36}$, corresponding to the best-fit value of $\alpha$, and $c_3=1$.

Applying the stability conditions constrains the model parameter to $0<c_2\leq 1$.
We then study the asymptotic behaviors:
\begin{align}
    f_{-}(R)/R_{0}
    &=
    -\frac{c_{2}R}{c_{3}R+R_{0}}
    \approx
    \begin{cases}
        -\frac{c_{2}}{c_{3}}+\frac{c_{2}}{c_{3}^{2}}\frac{R_{0}}{R}-\frac{c_{2}}{c_{3}^{3}}\left(\frac{R_{0}}{R}\right)^{2}\,,
        & R\gg R_{0}\,,\\
        -c_{2}\frac{R}{R_{0}}+c_{2}c_{3}\left(\frac{R}{R_{0}}\right)^{2}\,,
        & R\ll R_{0}\,,
    \end{cases}
\end{align}
\begin{align}
    f_{-}^{\prime}(R)
    &=
    -c_{2}\left(\frac{R_{0}}{c_{3}R+R_{0}}\right)^{2}
    \approx
    \begin{cases}
        -\frac{c_{2}}{c_{3}^{2}}\left(\frac{R_{0}}{R}\right)^{2}\,,
        & R\gg R_{0}\,,
        \\
        -c_{2}+2c_{2}c_{3}\frac{R}{R_{0}}-3c_{2}c_{3}^{2}\left(\frac{R}{R_{0}}\right)^{2}\,,
        & R\ll R_{0}\,,
    \end{cases}
\end{align}
and
\begin{align}
    R_{0}f_{-}^{\prime\prime}(R)
    &=
    2c_{2}c_{3}\left(\frac{R_{0}}{c_{3}R+R_{0}}\right)^{3}
    \approx
    \begin{cases}
        \mathcal{O}\left(\frac{1}{R^{3}}\right)
        \,, 
        & R\gg R_{0}
        \,,\\
        2c_{2}c_{3}\left[1-3c_{3}\frac{R}{R_{0}}+6c_{3}^{2}\left(\frac{R}{R_{0}}\right)^{2}\right]
        \,, 
        & R\ll R_{0}
        \,.
\end{cases}
\end{align}

\begin{figure}[htbp]
    \centering
    \includegraphics[scale=0.8]{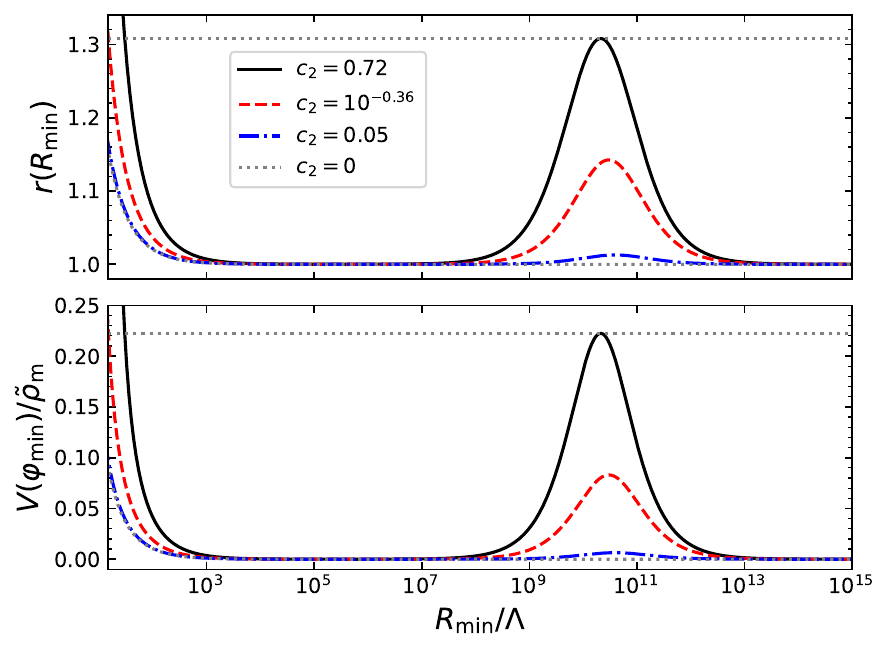}
    \caption{
        The upper panel shows $r(R_\mathrm{min})$ of model~\eqref{eq: ede mimus} as a function of $R_\mathrm{min}/\Lambda$, with $c_{3}=1$.
        Although we have fixed $R_{0}=4\times 10^{10}$, it only shifts the position of peaks in $r$ and does not affect the physics.
        The horizontal dotted line corresponds to $r=17/13$, required for $10\%$ energy injection.
        The lower panel shows the density ratio $V(\varphi_\mathrm{min})/\tilde{\rho}_\mathrm{m}$, with the same parameter choices as the upper panel. The horizontal dotted line corresponds to $V(\varphi_\mathrm{min})/\tilde{\rho}_\mathrm{m}=2/9$, required for $10\%$ energy injection.
        The growth of curves for $R/\Lambda<10^{3}$ indicates the onset of the DE-dominant epoch.
        }
    \label{fig: r_rho_minus}
\end{figure}

Substituting the model into Eq.~\eqref{eq: r} and imposing the condition $r(R_\mathrm{min})\geq 17/13$ yield
\begin{align}
    \frac{17c_{2}-8-\sqrt{289c_{2}^{2}-208c_{2}}}{8c_{3}}
    <
    \frac{R_{\mathrm{min}}}{R_{0}}
    \leq
    \frac{17c_{2}-8+\sqrt{289c_{2}^{2}-208c_{2}}}{8c_{3}}
    \,.
\end{align}
It has real solutions only if $c_{2}\geq\frac{208}{289}\simeq 0.72$, regardless of the specific value of $R_0$.
We find that the inferred value $c_{2}=10^{-0.36}\approx 0.44$ from Ref.~\cite{Odintsov:2023cli} is smaller than the lower bound.
Although the difference in $c_{2}$ is not large, the difference in fractional energy is quite significant.
As shown in Fig.~\ref{fig: r_rho_minus}, the EDE term can be nontrivial only if $c_{2}$ is on the order of one-tenth.
Using the best-fit value in Ref.~\cite{Odintsov:2023cli}, we find $V(c_2=10^{-0.36},c_3=1)/\tilde{\rho}_\mathrm{tot}\simeq 4\%$, corresponding to the dashed red line in Fig.~\ref{fig: r_rho_minus}.
We find that the best-fit value of $c_2$ is insufficient to achieve the required fractional energy.
However, $c_2=0.72$ corresponds to $\log \alpha \approx 7.86$, which is indeed within the range of 1$\sigma$ confidence level, and we can find $c_2 \gtrsim0.72$ consistent with the 1$\sigma$ region.

One observes that the current model can work well.
However, there exists a residual constant $c_{2}$ in $f_{R}(R)$ when $R\ll R_0$.
It is required that $c_{2}\ll 1$ to satisfy the constraint $F_R\simeq 1$ from the fifth-force experiments and observations, thereby restricting the cosmological impact of EDE.
We find that no admissible value of $c_2$ exists that can simultaneously explain the EDE and the local gravity tests, and, thus, this model is not viable.

To have a nontrivial EDE component leads to a large parameter $c_{2}$ appearing in $F_R(R)$ for $R\ll R_{0}$.
A consequence is that the EDE term would significantly change the late-Universe evolution if the late-time acceleration is also implemented in $F(R)$ gravity.
To see how it changes, we include a cosmological constant and the inflationary term in the $F(R)$ function:
\begin{align}
    F(R) = R - 2\Lambda + f_{\mathrm{ede}} + \frac{1}{6M^{2}}R^{2}
    \,,
    \label{eq: full F(R)}
\end{align}
where $M\simeq 1.5\times10^{-5}(50/N)M_{\mathrm{Pl}}$, with $N$ being the e-folding number during inflation.
The potential of the scalaron in the Einstein frame is given by
\begin{align}
    V(\varphi)/\rho_{\Lambda}&=\frac{1+\frac{R_{0}}{2\Lambda}\frac{c_{2}c_{3}R^{2}}{\left(c_{3}R+R_{0}\right)^{2}}+\frac{1}{6M^{2}}\frac{R^{2}}{2\Lambda}}{\left(1-\frac{c_{2}R_{0}^{2}}{\left(c_{3}R+R_{0}\right)^{2}}+\frac{R}{3M^{2}}\right)^{2}}
    \,,
    \label{eq: full potential minus}
\end{align}
where $\rho_{\Lambda}\equiv\frac{\Lambda}{\kappa^{2}}$.
In a well-constructed model, each term in the numerator and denominator should be relatively independent due to the gigantically different energy scales.
It is not true in our case, by looking at the asymptotic behavior:
\begin{align}
    V(\varphi)/\rho_{\Lambda}
    &\approx
    \begin{cases}
        \frac{R_{0}}{2\Lambda}
        \left[
        \frac{c_{2}}{c_{3}}-\frac{2c_{2}}{c_{3}^{2}}\frac{R_{0}}{R}+\frac{c_{2}\left(2c_{2}+3\right)}{c_{3}^{3}}\left(\frac{R_{0}}{R}\right)^{2}
        \right]\,, & R_{0}<R\ll M^{2}
        \,,\\
        \frac{1}{\left(1-c_{2}\right)^{2}}\left[1+\frac{4c_{2}c_{3}}{1-c_{2}}\frac{R}{R_{0}}+\left(c_{2}c_{3}\frac{R_{0}}{2\Lambda}+\frac{6\left(c_{2}^{2}+c_{2}\right)c_{3}^{2}}{\left(1-c_{2}\right)^{2}}\right)\left(\frac{R}{R_{0}}\right)^{2}\right]\,, & R\ll R_{0}\,.
    \end{cases}
\end{align}
For $R\ll R_{0}$, one observes that the factor $\frac{1}{\left(1-c_{2}\right)^{2}}>0$ coming from $f_\mathrm{ede}^{\prime}$ will cause a large potential $V/\rho_\Lambda>1$ even today.
To fit the observations, it indicates that the observed value should be an effective cosmological constant controlled by the EDE term:
\begin{align}
\label{eq: observed Lambda minus}
    \Lambda_\mathrm{obs}=\frac{\Lambda}{\left(1-c_{2}\right)^{2}}
    \,.
\end{align}
Note that $\Lambda$ is just a parameter in the $F(R)$ function.

\begin{figure}[htbp]
    \centering
    \includegraphics[scale=0.7]{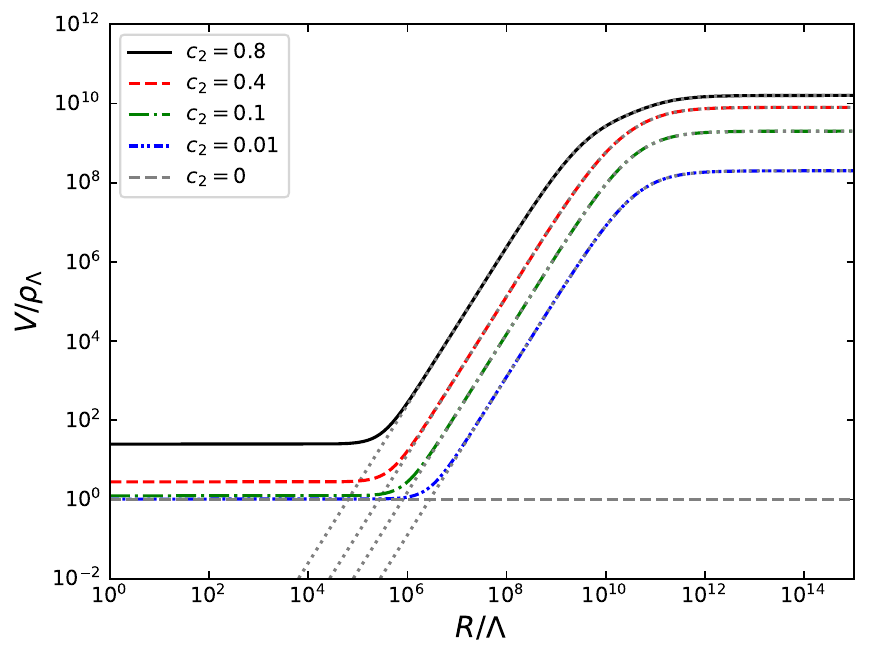}
    \caption{
        The Einstein-frame potential~\eqref{eq: full potential minus} of the scalaron versus the Ricci scalar $R/\Lambda$.
        In this figure, $c_{3}=1$ and $c_{2}=0-0.8$ from bottom to top.
        The horizontal dashed line corresponds to the cosmological constant $\Lambda$.
        The dotted gray lines show how the EDE decays without a cosmological constant in the $F(R)$ function.
        With the presence of a cosmological constant, on the other hand, the EDE term reshapes an effective cosmological constant, which should be the observed one~\eqref{eq: observed Lambda minus}.
        }
    \label{fig: Vminus_comparison}
\end{figure}

\begin{figure}[htbp]
    \centering
    \includegraphics[scale=0.7]{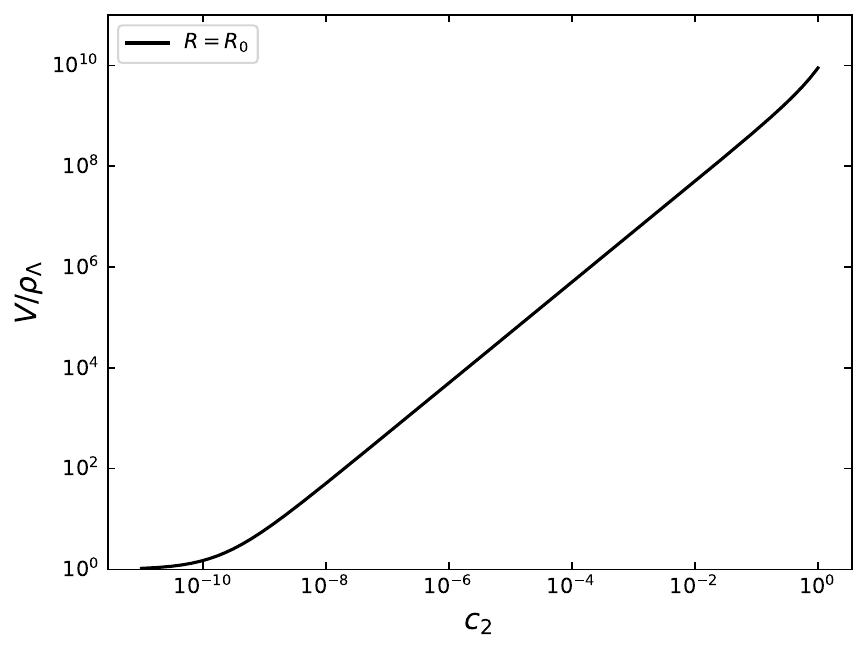}
    \caption{
        The Einstein-frame potential value~\eqref{eq: full potential minus} of the scalaron versus parameter $c_{2}$ for $R=R_{0}$ and $c_{3}=1$.
        }
    \label{fig: Vminus_c2}
\end{figure}

\begin{figure}[htbp]
    \centering
    \includegraphics[scale=0.7]{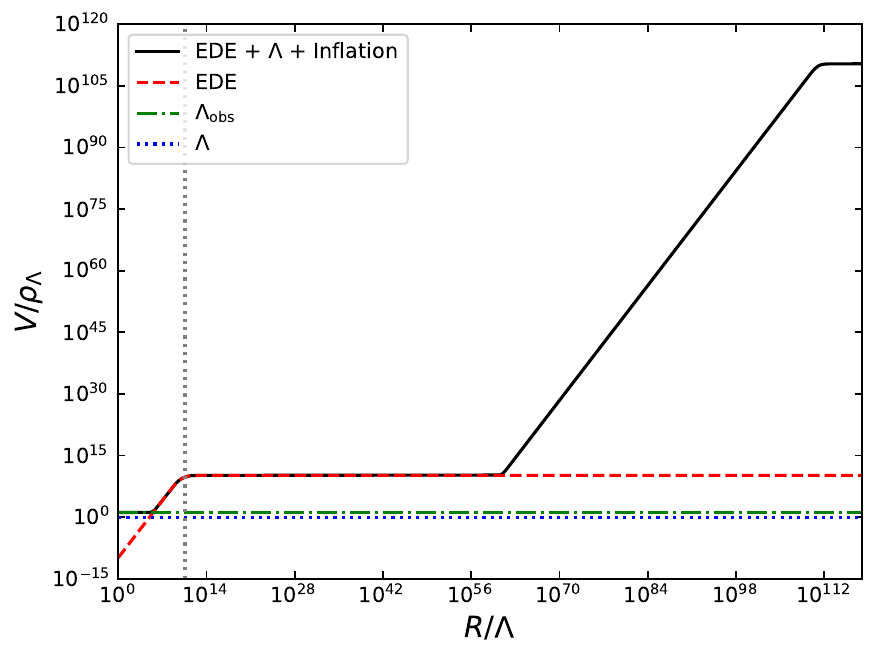}
    \caption{
        The Einstein-frame potential~\eqref{eq: full potential minus} of the scalaron versus the Ricci scalar,
        with $c_2=0.72$ and $c_3=1$.
        The EDE term corresponds to model~\eqref{eq: ede mimus}, $\Lambda$ is the cosmological constant (or parameter) appearing in the $F(R)$ function, and $\Lambda_\mathrm{obs}$ is the observed cosmological constant given in Eq.~\eqref{eq: observed Lambda minus}.
        The vertical dotted line corresponds to $R=R_{0}$.
        }
    \label{fig: potential minus}
\end{figure}

Figure~\ref{fig: Vminus_comparison} shows the influence of the parameter $c_{2}$ on the intermediate- and late-time de Sitter points.
Figure~\ref{fig: Vminus_c2} shows the height of the saddle $V(R=R_{0})$ as a function of the parameter $c_{2}$.
The EDE term soon becomes negligible if $c_{2}\ll 1$, which does not rely on the specific value of $R_{0}$.
Figure~\ref{fig: potential minus} shows three saddles in the potential, which provides us with a straightforward comparison of how different terms are related.
It is easy to extend our discussion to more complicated DE terms in $F(R)$ gravity, whose parameter values should also be modified by the EDE term.
On the other hand, without a DE term in $F(R)$ gravity, the EDE term would decay quickly when $R\ll R_{0}$,
as shown by dotted lines in Fig.~\ref{fig: Vminus_comparison}.
Therefore, this saddle EDE model is compatible with other DE models from different theories.

So far, everything seems to work well in this model, especially since we have absorbed the residual constant $c_{2}$ into the parameter $\Lambda$ to achieve successful late-time acceleration.
However, one recognizes that $F_{R}\simeq 1-c_{2}$ for $R<R_{0}$ still plague the model.
The current model of $F(R)$ gravity should mimic the $\Lambda$CDM model after the matter-dominated epoch and satisfy the stringent bound on $F_{R}$ from the tests of the violation of the equivalence principle, which requires $F_R\simeq 1$~\cite{Capozziello:2007eu}.
It is, therefore, required that $c_{2}\ll 1$ for the EDE to be cosmologically viable, although, in this case, it does not bring any new recombination physics.
Precisely, the upper limit of the stabilized scalaron field for critical galaxy density, $\varphi_\mathrm{galaxy}/M_\mathrm{Pl}<10^{-15}$~\cite{Capozziello:2007eu}, constrains $c_{2}<10^{-15}$, which implies that the EDE term is negligible during the cosmic time until today, as shown in Figs.~\ref{fig: Vminus_comparison} and~\ref{fig: Vminus_c2}.

Lastly, we propose a similar model to $f_{-}(R)$:
\begin{align}
    f_{\mathrm{ede}} (R)
    = 
    -c_{2}R_{0}\frac{Re^{R/R_{0}}}{c_{3}Re^{R/R_{0}}+R_{0}}
    \,.
\end{align}
The required parameter value $c_2=0.53$ of this model is smaller than that of the negative model~\eqref{eq: ede mimus}.
It also exhibits faster asymptotic behavior approaching a constant in the large-curvature limit.
However, looking at $F_R(R)$ for $R\ll R_{0}$, we find
\begin{align}
    F_{R}(R)
    \approx
    1-c_{2}
    -2c_{2}\left(1-c_{3}\right)\frac{R}{R_{0}}
    -3c_{2}\left(c_{3}^{2}-2c_{2}+\frac{1}{2}\right)\left(\frac{R}{R_{2}}\right)^{2}
    \,.
\end{align}
Thus, this model suffers from the same problem that $F_R$ is away from unity due to the nonvanishing constant $c_{2}$.

%%%%%%%%%%%%%%%%%%%%%%%%%
\subsubsection{Case 2: \texorpdfstring{$f_{\mathrm{ede}}>0$}{TEXT}}

Finally, we turn to the positive EDE~\eqref{eq: ede plus} with a positive sign.
We will show that the situation is more difficult than the former one.
This model satisfies all stability conditions, $F, F_{R}, F_{RR}>0$, as long as $c_{1},c_{3}>0$.
Therefore, $c_{1}$ is less constrained compared with $c_{2}$ in model~\eqref{eq: ede mimus}.
We then study the asymptotic behaviors:
\begin{align}
    f_{+}/R_{0}
    &=
    \frac{c_{1}R^{2}}{c_{3}R_{0}R+R_{0}^{2}}\approx
    \begin{cases}
        \frac{c_{1}}{c_{3}}\frac{R}{R_{0}}-\frac{c_{1}}{c_{3}^{2}}\left(1-\frac{R_{0}}{c_{3}R}+\left(\frac{R_{0}}{c_{3}R}\right)^{2}\right)
        \,, 
        & R\gg R_{0}
        \,,\\
        c_{1}\left(\frac{R}{R_{0}}\right)^{2}
        \,, 
        & R\ll R_{0}
        \,,
    \end{cases}
\end{align}
\begin{align}
    f_{+}^{\prime}
    &=
    \frac{c_{1}R\left(c_{3}R+2R_{0}\right)}{\left(c_{3}R+R_{0}\right)^{2}}\approx
    \begin{cases}
        \frac{c_{1}}{c_{3}}\left(1-\left(\frac{R_{0}}{c_{3}R}\right)^{2}\right)
        \,, 
        & R\gg R_{0}
        \,,\\
        2c_{1}\frac{R}{R_{0}}-3c_{1}c_{3}\left(\frac{R}{R_{0}}\right)^{2}
        \,, 
        & R\ll R_{0}
        \,,
    \end{cases}
\end{align}
and
\begin{align}
    R_{0}f_{+}^{\prime\prime}
    &=
    \frac{2c_{1}R_{0}^{3}}{\left(c_{3}R+R_{0}\right)^{3}}
    \approx
    \begin{cases}
        \mathcal{O}\left(\frac{1}{R^{3}}\right)
        \,, 
        & R\gg R_{0}
        \,,\\
        2c_{1}\left(1-3c_{3}\frac{R}{R_{0}}+6c_{3}^{2}\left(\frac{R}{R_{0}}\right)^{2}\right)
        \,, 
        & R\ll R_{0}
        \,.
    \end{cases}
\end{align}
Again, one observed that $f_{+}$ and $f_{+}^{\prime\prime}$ are well behaved; however, there exists a residual constant $c_{1}/c_{3}$ in $F_R$ for $R\gg R_0$, indicating that $c_{1}/c_{3}\ll 1$.
Nevertheless, we continue to study the necessary conditions to alleviate the Hubble tension.
Imposing the condition $r\geq 17/13$ on the model gives
\begin{align}
    \frac{9c_{1}-8c_{3}-\sqrt{\left(8c_{3}-9c_{1}\right)^{2}-64c_{3}\left(c_{1}+c_{3}\right)}}{8c_{3}\left(c_{1}+c_{3}\right)}
    <
    \frac{R_{\mathrm{min}}}{R_{0}}
    <
    \frac{9c_{1}-8c_{3}+\sqrt{81c_{1}^{2}-208c_{1}c_{3}}}{8c_{3}\left(c_{1}+c_{3}\right)}
    \,.
\end{align}
It has real solutions only if $c_{1}/c_{3}\geq 208/81\gtrsim 2.57$.

\begin{figure}[htbp]
    \centering
    \includegraphics[scale=0.7]{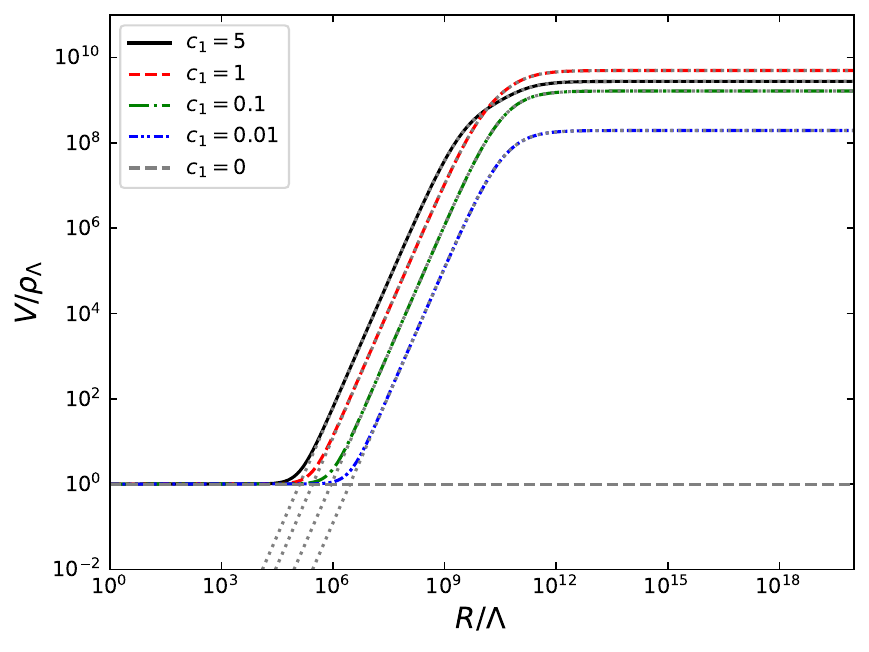}
    \caption{
        The Einstein-frame potential~\eqref{eq: full Vplus} of the scalaron versus the Ricci scalar $R/\Lambda$.
        In this figure, $c_{1}=0$--$5$ and $c_{3}=1$.
        The horizontal dashed line corresponds to the cosmological constant $\Lambda$.
        The dotted gray lines show how the EDE decays without a cosmological constant in the $F(R)$ function, indicating that the EDE term does not affect the late-time DE epoch.
        }
    \label{fig: Vplus_comparison}
\end{figure}

\begin{figure}[htbp]
    \centering
    \includegraphics[scale=0.7]{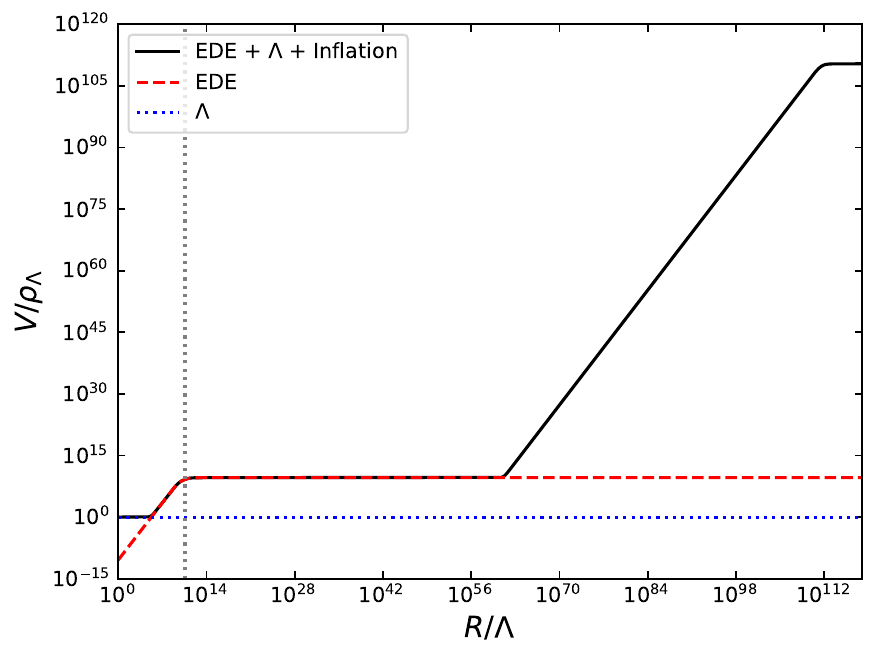}
    \caption{
        The Einstein-frame potential~\eqref{eq: full Vplus} of the scalaron versus the Ricci scalar,
        with $c_{1}=2.6$ and $c_{3}=1$.
        The EDE term corresponds to model~\eqref{eq: ede plus}, $\Lambda$ is the cosmological constant, and the vertical dotted line corresponds to $R=R_{0}$.
        }
    \label{fig: Vplus}
\end{figure}

Considering the full functional form of $F(R)$ gravity~\eqref{eq: full F(R)}, the Einstein-frame potential is written as
\begin{align}
    V(\varphi)/\rho_{\Lambda}
    &=
    \frac{1+\frac{c_{1}R^{2}}{\left(c_{3}R+R_{0}\right)^{2}}\frac{R_{0}}{2\Lambda}+\frac{1}{6M^{2}}\frac{R^{2}}{2\Lambda}}{\left(1+\frac{c_{1}R\left(c_{3}R+2R_{0}\right)}{\left(c_{3}R+R_{0}\right)^{2}}+\frac{1}{3M^{2}}R\right)^{2}}
    \,.
    \label{eq: full Vplus}
\end{align}
It has the following asymptotic behaviors:
\begin{align}
    V(\varphi)/\rho_{\Lambda}
    &\approx
    \begin{cases}
        \frac{R_{0}}{2\Lambda}\left[\frac{c_{1}}{\left(c_{1}+c_{3}\right)^{2}}\left(1-\frac{2R_{0}}{c_{3}R}\right)+\frac{c_{1}\left(5c_{1}+3c_{3}\right)}{c_{3}^{2}\left(c_{1}+c_{3}\right)^{3}}\left(\frac{R_{0}}{R}\right)^{2}\right]
        \,, 
        & R_{0}<R\ll M^{2}
        \,,\\
        1-4c_{1}\frac{R}{R_{0}}+\left(12c_{1}^{2}+6c_{3}c_{1}+c_{1}\right)\left(\frac{R}{R_{0}}\right)^{2}
        \,, 
        & R\ll R_{0}
        \,.
    \end{cases}
\end{align}
Unlike the former case, although this model also has a constant in the potential, it does not change the low-curvature term when we implement the late-time acceleration in $F(R)$ gravity, as shown in Figs.~\ref{fig: Vplus_comparison} and~\ref{fig: Vplus}.
In this sense, the positive EDE works better than the negative EDE.

\begin{figure}[htbp]
    \centering
    \includegraphics[scale=0.7]{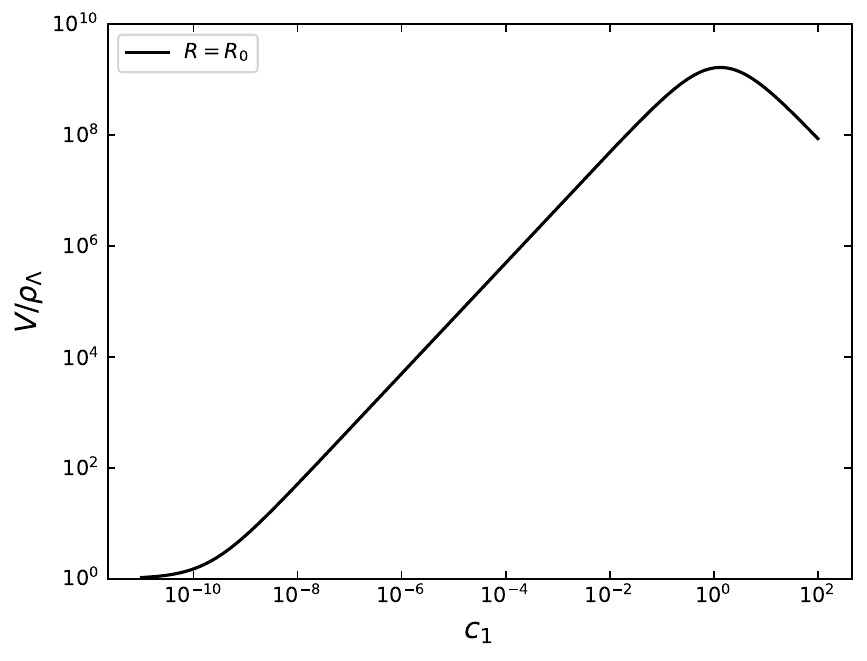}
    \caption{
        The Einstein-frame potential value~\eqref{eq: full Vplus} of the scalaron versus parameter $c_{1}$ for $R=R_{0}$ and $c_{3}=1$.
        }
    \label{fig: Vplus_c1}
\end{figure}

Returning to the constraints on $F_R(R)$, however, we find that the problem still exists for $R>R_{0}$.
The residual constant $c_{1}/c_{3}\gtrsim 2.57$ rules out this model as a viable cosmological framework.
On the one hand, it violates the BBN constraints $F_{R}(R_{\mathrm{BBN}})\lesssim\left(\frac{10}{9}\right)^{2}$ [see Eq.~\eqref{eq: BBN bound}].
On the other hand, it is much larger than the constraints inferred from the local tests of violation of the equivalence principle, i.e., $F_{R}(\varphi_\mathrm{atm})\simeq 1$~\cite{Capozziello:2007eu}.
Precisely speaking, $\varphi_\mathrm{atm}/M_\mathrm{Pl}<10^{-15}$ constrains that $c_{1}/c_{3}<10^{-15}$, implying a negligible contribution of the EDE term in the cosmic history, as shown in Figs.~\ref{fig: Vplus_comparison} and~\ref{fig: Vplus_c1}.

In summary, neither of the two saddle models~\eqref{eq: ede mimus} and \eqref{eq: ede plus}, negative and positive EDE models, can have any appreciable impact on cosmic history while passing the local gravity tests.
They certainly cannot provide the required energy injection at MRE and introduce sufficiently large EDE to alleviate the Hubble tension once the constraints from fifth-force observations are taken into account.

%%%%%%%%%%%%%%%%%%%%%%%%%
%%%%%%%%%%%%%%%%%%%%%%%%%
\section{Constraints on EDE models}
\label{No-go theorem}

%%%%%%%%%%%%%%%%%%%%%%%%%
\subsection{Generic constraint on the potential-driven EDE}

In the previous section, we investigated possible scenarios of EDE in $F(R)$ gravity and found that all failed to pass the cosmological constraints from BBN and/or the experimental bounds from the violation of the equivalence principle.
Although it is not a scrutiny for all possible scenarios, the difficulty of developing EDE in $F(R)$ gravity leads us to reconsider the relationship among EDE, $F_{R}$ (or $\varphi$), and experimental constraints.

We first assumed the stability conditions for the $F(R)$ function based on theoretical considerations.
Based on the potential-driven EDE scenario, we have computed the conditions under which the scalaron potential energy is sufficiently large to alleviate the Hubble tension.
Adopting observational constraints from local gravity tests, we find that they are incompatible with resolving the Hubble tension.
Independent of the details of models, the above consideration points to a generic constraint on the potential-driven EDE scenarios in $F(R)$ gravity.

\begin{figure}[htbp]
    \centering
    \includegraphics[scale=0.7]{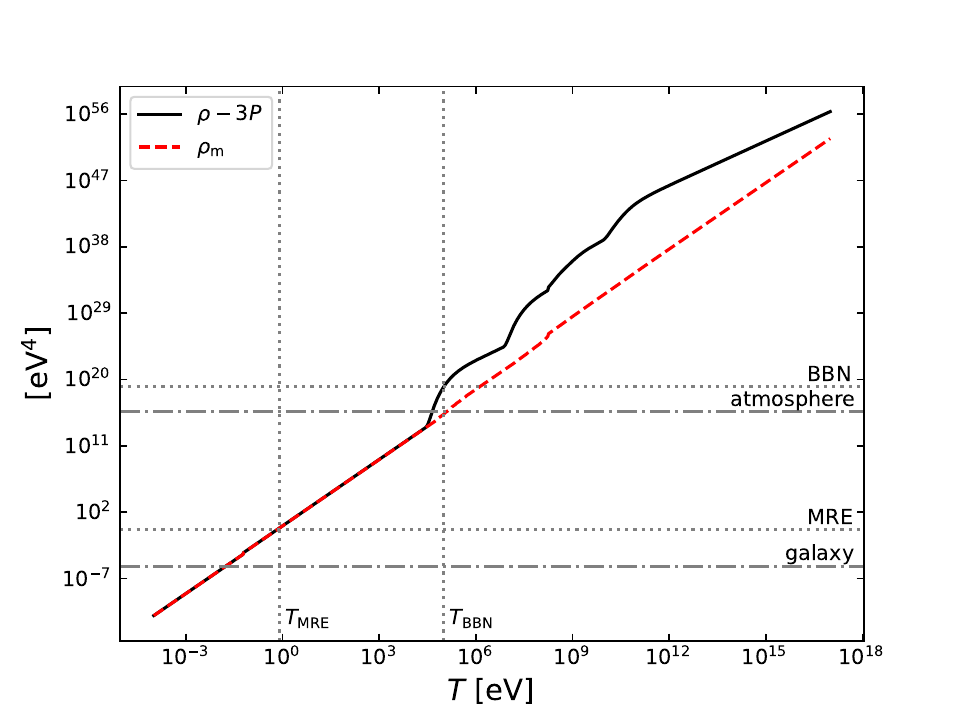}
    \caption{
        The cosmic background evolution of matter density (red dashed line) and the trace of energy-momentum tensor (black solid line) as a function of Jordan-frame temperature.
        The two vertical dotted lines denote the moment of MRE and BBN, respectively.
        The two horizontal dotted lines denote the values of the trace of the energy-momentum tensor for MRE and BBN, respectively.
        For comparison, the two horizontal dot-dashed lines correspond to the matter density in the galaxy and the atmosphere, respectively.
        One observes that $\rho_\mathrm{galaxy}\ll\rho_\mathrm{m}(T_\mathrm{MRE})\ll\rho_\mathrm{atm}\sim\rho_\mathrm{m}(T_\mathrm{BBN})\ll -T_\mu^{\mu}(T_\mathrm{BBN})$.
        }
    \label{fig: cosmic_and_local_density}
\end{figure}

Generalizing our results in the previous section, we find the following consequences:
\begin{enumerate}[label=(\roman*), leftmargin=3em, itemsep=0.5em]
\item \textit{Theoretical constraints.} 
For stability, the $F(R)$ gravity theory requires $F(R)>0$ [see Eq.~\eqref{eq: stability condition}], $F_{R}(R)>0$ (to avoid antigravity), and $F_{RR}(R)>0$ (to avoid tachyonic
instability) for $R>R_{0}$, where de Sitter solution is realized at $R=R_{0}$ to describe EDE.
Consequently, $F(R)$ is a convex and monotonically increasing function of $R$, and $F_{R}(R)$ is a monotonically increasing function of $R$.
The Jordan-frame scalaron field $\Phi\equiv F_{R}(R)$ or the Einstein-frame scalaron field $\varphi\equiv \frac{\sqrt{6}M_\mathrm{Pl}}{2}\ln F_{R}(R)$ is also a monotonically increasing function of $R$.

\item \textit{Experimental constraints.}
The local tests of the violation of the equivalence principle have placed stringent constraints on the field amplitude $\varphi_{\mathrm{min}}$ responsible for matter density ranging from the atmosphere to the galaxy~\cite{Capozziello:2007eu}.
The experimental bound is $\left|\varphi_\mathrm{atm}\right|<\left|\varphi_\mathrm{galaxy}\right|<10^{-15}M_\mathrm{Pl}$ or, for $F(R) = R + f(R)$, $f_{R}(R)\lesssim 10^{-15}$.
This indicates that $f(R)$ is a slowly evolving function.
Since $f_{RR}=F_{RR}>0$, $f_{R}(R)$ is a slowly increasing function.

\item \textit{Constraints on $r$.} 
The lower bound $r(R_{\mathrm{min}})\geq1$ leads to
\begin{equation}
f_{R}(R_{\mathrm{min}})\geq\frac{f(R_{\mathrm{min}})}{R_{\mathrm{min}}}
\,.
\end{equation}
The constraint $f_R\lesssim 10^{-15}$ together with $F>0$ implies that $-1<\frac{f(R_{\mathrm{min}})}{R_{\mathrm{min}}}<10^{-15}$.
We discuss $f>0$ and $f<0$ separately below.
\begin{itemize}
\item If $f>0$, then we have $0<\frac{f}{R}<10^{-15}$ and arrive at
\begin{equation}
\begin{aligned}
r(R_{\mathrm{min}})&=\frac{1+f_{R}(R_{\mathrm{min}})}{1+f(R_{\mathrm{min}})/R_{\mathrm{min}}}\\&\approx1+f_{R}(R_{\mathrm{min}})-f(R_{\mathrm{min}})/R_{\mathrm{min}}\\&\lesssim1+f_{R}(R_{\mathrm{min}})
\,.
\end{aligned}
\end{equation}
Therefore, $0\leq r(R_{\mathrm{min}})-1\lesssim 10^{-15}\ll 1$.

\item For the case that $f<0$, it is a little bit complicated. 
Since $f(R)$ is very slowly changing, we can expand it as $f(R)\equiv-\Lambda_{\mathrm{e}}+\varepsilon(R)$, where $\Lambda_{\mathrm{e}}$ is a constant, $|\varepsilon(R)|\ll R$, $|\varepsilon_{R}(R)|\ll1$, and $\varepsilon_{RR}(R)>0$.
Note that this approximation is valid for a large range of Ricci scalar, $R_{\mathrm{galaxy}}<R<R_{\mathrm{atm}}$, corresponding to matter density in the regime $\rho_{\mathrm{galaxy}}<\rho_{\mathrm{m}}<\rho_{\mathrm{atm}}$.
Therefore, the constraint $F=R-\Lambda_{\mathrm{e}}+\varepsilon(R)>0$ implies that $\Lambda_{\mathrm{e}}\lesssim R_\mathrm{galaxy}\ll R_\mathrm{atm}$.
Then we find
\begin{align}
    r(R_{\mathrm{min}})
    =
    \frac{1+f_{R}(R_{\mathrm{min}})}
    {1-\Lambda_{\mathrm{e}}/R_{\mathrm{min}}+\varepsilon(R)/R_{\mathrm{min}}}
    \,.
\end{align}
For large curvature $R\gg\Lambda_\mathrm{e}$,
\begin{equation}
    r(R_{\mathrm{min}})
    \approx
    1+f_{R}(R_{\mathrm{min}})
    +\frac{\Lambda_{\mathrm{e}}}{R_{\mathrm{min}}}
    -\frac{\varepsilon(R)}{R_{\mathrm{min}}}
    \,,
\end{equation}
which again indicates $r(R_{\mathrm{min}})-1\ll 1$.
\end{itemize}
So, inferred from local gravity tests, it proves that $r(R_{\mathrm{min}})-1\ll 1$ in the density region $\rho_{\mathrm{galaxy}}<\rho_{\mathrm{m}}<\rho_{\mathrm{atm}}$.

\item \textit{Constraints on potential.}
$r(R_{\mathrm{min}})-1\ll 1$ implies that the potential should be steep. 
To see this, we define a new dimensionless quantity, analogous to the slow-roll parameter in inflation, to estimate the flatness of the potential:
\begin{equation}
    \epsilon
    \equiv
    \frac{M_{\mathrm{Pl}}^{2}}{2}\left(\frac{V_{,\varphi}}{V}\right)^{2}
    =
    \frac{1}{3}\left(\frac{2-r(R_{\mathrm{min}})}{r(R_{\mathrm{min}})-1}\right)^{2}
    \,,
\end{equation}
where we have imposed an exponent $2$ to make $\epsilon$ positive-defined, although it is not necessary for the current consideration because $V_{,\varphi}, V>0$ in $F(R)$ gravity for $R>R_{1}$.
One observes that $r(R_{\mathrm{min}})-1\ll1$ leads to $\epsilon\gg1$, indicating a steep potential.
On the other hand, $\epsilon\ll1$ only if $r\simeq 2$. It implies the possible presence of a de Sitter phase associated with a flat potential. This is the theoretical basis of DE scenarios in $F(R)$ gravity.

\item \textit{Impacts on cosmology.}
The above results are derived from local gravity tests that omit scalaron dynamics. 
Assuming the scalaron was stabilized in the effective potential in the target cosmic history, we can apply the above results to cosmology, where the scalaron dynamics could be nontrivial.
Considering the equation-of-state parameter $w=P/\rho$,
one finds $\rho_\varphi=\frac{2V}{1-w}$.
Then, we obtain 
\begin{align}
    \rho_\varphi(\varphi_\mathrm{min})/\rho_\mathrm{m}=\frac{2}{1-w}V(\varphi_\mathrm{min})/\rho_\mathrm{m}\ll 1
\end{align} 
for $w<1$.

In other words, the scalaron energy density is always negligible compared with matter density, so its cosmological impact is insignificant compared with ordinary matter (at least, the classical level).
This statement is general no matter what role the scalaron played in cosmic history, such as radiation $w\approx 1/3$, dust $w\approx 0$, or early dark energy $w\approx -1$, as long as the value of matter density in the target cosmic period ranges from $\rho_{\mathrm{m}}(T)=\rho_{\mathrm{atm}}$ to $\rho_{\mathrm{m}}(T)=\rho_{\mathrm{galaxy}}$ (see Fig.~\ref{fig: cosmic_and_local_density}).
\end{enumerate}

There are two key points to be emphasized.
First, theoretical requirements on the $F(R)$ function for the stability indicate that $F_R(R)$ is an increasing function of the curvature.
Second, as in Fig.~\ref{fig: cosmic_and_local_density}, the background curvature at the MRE epoch is between that of the BBN epoch or atmosphere and that of the galaxy. 
As a result, the value of $F_R$ at MRE is sandwiched by two independent constraints, and it is inevitable to encounter the constraint $F_R\approx1$ at MRE, which suppresses the cosmological impact of the potential energy of the scalaron as EDE.

The above statement is better understood by comparing potential-driven EDE with the LDE models in $F(R)$ gravity.
For illustration, let us consider $F(R)=R-2\Lambda$, where the cosmological constant sources the LDE.
This model always leads to $F_{R}=1$ and $r(R_\mathrm{min})=\frac{R}{R-2\Lambda}$.
One finds that $r(R_\mathrm{LDE})\sim 31/17$ in the current LDE-dominated Universe, which contradicts our statement that $F_{R}(R)$ points to $r\approx 1$.
The discrepancy arises because, at present, the Ricci scalar has decreased to a value comparable to $2\Lambda$, which significantly reduces $F(R)$ in the denominator of $r$.
By contrast, in the potential-driven EDE scenario, we consider $F(R)=R+f_\mathrm{ede}(R)$, where $f_\mathrm{ede}(R)$ is characterized by a critical curvature $R_{0}$.
The Universe evolves from a high-curvature regime $R\gg R_{0}$ to a low-curvature regime $R\ll R_{0}$, and $F(R)$ remains a convex, monotonically increasing function throughout this history. 
This structure implies that $f(R)\ll R$ whenever $\left|f_{R}(R)\right|=\left|F_{R}(R)-1\right|\ll1$ [see step 4], even if the functional form of $f(R)$ is complicated.
Consequently, $r\approx 1$ holds for the potential-driven EDE realization within $F(R)$ gravity.

It is also worth noting that $|\varphi|\ll 1$ in the early Universe, indicating that our analysis is essentially frame independent.
Furthermore, our assumption regarding the potential-driven EDE mechanism is physically reasonable.
As shown in Ref.~\cite{Chen:2022zkc}, although the scalaron initially remained frozen far from the effective potential minimum due to strong Hubble friction, the decoupling of SM particles from the thermal bath deep in the radiation-dominated era induces an effective external force that drives the scalaron toward its minimum.

%%%%%%%%%%%%%%%%%%%%%%%%%
\subsection{Time-varying gravitational constant in the early Universe}

We have demonstrated that, at the background level, the potential-driven EDE scenario is challenging to realize in $F(R)$ gravity while satisfying existing observational constraints from local gravity tests.
The essential problem lies in the value of $F_{R}(R)$, and it is stringently constrained by independent observations on different curvature scales.
We can apply the previous consideration in the potential-driven EDE to a more general situation.
That is, regardless of the feasibility of EDE, constraints on $F_{R}(R)$ restrict the modification to GR in the early Universe.
For the power-law model, $F_R$ is not prevented from tending to infinity if we fix the mass scale to explain the period of MRE. 
For the base saddle models, there is a residual constant in $F_R$, which diminishes the early-Universe effect of the scalaron to pass the local gravity tests.
Therefore, it is simply hard to change $F_{R}$ in the early Universe when we demand consistency with the results in local observations of $F_{R}$.

Following the potential-driven EDE scenario, we can further discuss the generic constraints on early-Universe models of $F(R)$ gravity.
Assuming that the scalaron field follows from the stationary condition, we summarize the conditions for the $F(R)$ functions and the consequences of adopting the local gravity tests.
\begin{enumerate}
    \item We impose the stability conditions $F(R), F_{R}(R), F_{RR}(R)>0$.
    Consequently, $F(R)$ is a convex and monotonically increasing function, and $F_{R}(R)$ is a monotonically increasing function.

    \item
    We do not consider the potential-driven EDE but impose the quasistatic background evolution determined by the effective potential minimum and matter-energy density.
    This assumption enables us to adopt the observational constraints from local gravity tests, which are also applicable in the stationary condition.
    
    \item 
    We adopt the experimental and observational bounds.
    We consider not only the background energy density at MRE but also a wide range of the matter-energy density $\rho_\mathrm{m}$ in the early Universe.
    We adopt the constraints at $\rho_\mathrm{galaxy}$ and $\rho_\mathrm{BBN}$, and, thus,
    $\rho_\mathrm{galaxy}<\rho_\mathrm{m}<\rho_\mathrm{BBN}$.
    
    \item 
    $\left|\varphi_\mathrm{galaxy}\right| \ll M_\mathrm{Pl}$ and $\left|F_{R}(R_\mathrm{galaxy})-1\right|\ll 1$.
    A conservative theoretical bound applied to the BBN epoch gives $(10/11)^2\lesssim F_{R}(R_\mathrm{BBN})\lesssim (10/9)^2$, and, thus, $\left|F_{R}(R_\mathrm{BBN})-1\right|\ll 1$.
    Because $F_{R}(R)$ is the increasing function of $R$,
    we obtain $F_R(R_\mathrm{galaxy}) -1 <F_R(R_\mathrm{m}) -1< F_R(R_\mathrm{BBN}) -1 \ll 1$.
    Thus, we obtain the constraint $F_R(R) \approx 1$ from $1$ MeV (temperature at BBN epoch) to $10^{-2}$ eV (corresponding to galaxy-scale curvature) in the energy scale.
\end{enumerate}

It is worth mentioning that $F_R$ corresponds to the time-varying gravitational constant in the $F(R)$ gravity.
The modified gravity theories generally predict such a running Planck mass, and several works have considered the decaying or transitional Planck mass for the Hubble tension
~\cite{Lima:2016npg, Ballardini:2020iws, Ballardini:2021evv, Ballesteros:2020sik, Braglia:2020auw, Zumalacarregui:2020cjh, Benevento:2022cql, Kable:2023bsg}.
As we have demonstrated, local observational constraints and the increasing $F_R$ behavior with respect to $R$ suggest that $F_R$ should be nearly unity in the early Universe.
Thus, we can conclude that large changes in the gravitational constant in the early Universe can be constrained.

Note that the above discussion assumes the scalaron field remains at its potential minimum during time evolution, ignoring its kinetic energy and rapid changes in its field value.
The time evolution of $F_R$ is related to $F_{RR}$ as $dF_R / dt = dR/dt \cdot F_{RR}$.
Because $dR/dt$ is determined by the cosmic history, the slowly evolving gravitational constant generally indicates a small $F_{RR}$.
In other words, if we demand that $F_{RR}$ is small, it would be hard to realize the significant change of the effective gravitational constant.
It is notable that the effective mass of the scalaron is proportional to the inverse of $F_{RR}$ as in Eq.~\eqref{m2} and that the small $F_{RR}$ indicates the large effective mass.
The experimental constraint on $F_{RR}$ comes from the upper bound on the scalaron mass from the fifth-force search experiment, and we did not account for this constraint in our current analysis.
Thus, it would be possible to weaken the condition for $F_{RR}$, and the transitional Planck mass scenario can be realized in such a setup.
We will reconsider the constraints on scalaron mass or $F_{RR}$ in the conclusion.

%%%%%%%%%%%%%%%%%%%%%%%%%
%%%%%%%%%%%%%%%%%%%%%%%%%
\section{Conclusion}
\label{Conclusion}

This paper has investigated several possible EDE models in $F(R)$ gravity.
In particular, we have used these models as a recombination solution to the Hubble tension problem, in the same manner as the existing EDE scenarios, by imposing a $10\%$ fraction of the energy density at the MRE period.
To achieve this purpose, it is required that the density ratio of the scalaron and matter (dust and dark matter) should be no less than $2/9$ at MRE.
We, therefore, constructed a dimensionless quantity $r(R_\mathrm{min})\equiv R_\mathrm{min}F_{R}(R_\mathrm{min})/F(R_\mathrm{min})$ in Eq.~\eqref{eq: r} to quantify the density ratio~\eqref{eq: density ratio}.
The advantage of using this quantity is that, firstly, it can be calculated directly from the $F(R)$ function, and, secondly, it is also a bounded quantity $1\leq r \leq 2$; especially, $r\approx 1$ required by a matter-dominated epoch, $r\approx 31/17$ today, $r=2$ at the (stable) de Sitter point, and $r\geq 17/13$ for enough energy injection at MRE.

We then studied the asymptotic behaviors of these models. We have demonstrated that power-law models $F(R)\propto R^{n}$ are undesired, since they would cause $F_{R}\gg 1$ for $R\gg m^2$. Regarding saddle EDE models that shape a saddle in the scalaron potential $V(R(\varphi))$,
we considered two base models and found that saddle functions generally produce an unwanted residual constant in the $F_{R}$ function at the large- or small-curvature limit, which significantly limits the cosmological effects of such models.
On the one hand, to pass local gravity tests and satisfy the BBN bound, the constant should be much less than unity.
On the other hand, to make the scalaron contribute a nontrivial fractional energy density in the intermediate epoch, the constant should be on the order of unity.
This dilemma leads to a breakdown of $F(R)$ gravity as a remedy for the Hubble tension problem, in the same way as in ordinary EDE scenarios.

We found that the problem lies mainly in $F_{R}(R)$, leading us to reconsider the correlation between local gravity tests and cosmic background evolution, both of which are linked by the quasistatic evolution of the scalaron field.
By summarizing the necessary conditions for a viable $F(R)$ model and assuming that the scalaron always followed the effective potential minimum, we eventually face the general constraint on $F(R)$ gravity:
If there exists an EDE in cosmic history with matter density ranging from $\rho_\mathrm{galaxy}<\rho<\rho_\mathrm{atm}$, the energy density of EDE, the scalaron field in our analysis, is negligible compared with matter density and can be neglected in the cosmic evolution.

In closing, we make several remarks on our results.
First, we revisit the validity of the assumptions we have imposed and the conclusions we draw from them.
We have considered the potential-driven EDE by ignoring the kinetic energy of the scalaron.
There is room for nontrivial kinetic-energy contributions to EDE to address the Hubble tension, which is indeed outside our proposed constraint from the minimal setup.
At the same time, if we take into account the kinetic energy, we cannot adopt the observational constraint from the local gravity test, since the time-dependent scalaron field value need not always satisfy the constraints derived for a static background.
That is, the value of $F_R$ \textit{at a moment in the early Universe} can violate the constraints from the local tests, because there is no quasistatic evolution.
Moreover, if the scalaron does not follow the evolution of the stationary condition, it is inevitable to study the independent and nonperturbative dynamics of the scalaron.
For instance, in this work, we did not account for perturbations of the scalaron.
As several existing works have shown, it is necessary to solve the perturbative scalaron equation on the cosmological background and to analyze the influence on the CMB spectrum.

Second, we reconsider the fluctuation and perturbation around the potential minimum.
As briefly explained in the last subsection, we did not utilize any constraints on $F_{RR}$.
$F_{RR}$ is related not only to the time evolution of the gravitational constant,
but also to the scalaron effective mass $m^2\propto 1/(3F_{RR}(R))$.
Viable models of $F(R)$ gravity should exhibit the chameleon mechanism to hide the fifth force in local gravity tests.
Since we could have applied the late-time observational constraints to the early-Universe model, we expect the chameleon mechanism to also work in the early Universe.
In this case, the scalaron mass is expected to be larger than the energy density of the other matter contents, and, thus, the fluctuating mode of the scalaron is also expected to be suppressed.
Therefore, it would still be challenging to invoke the nontrivial perturbative contribution to the Hubble tension.
However, outside the potential-driven scenario, there is room to search for EDE models of $F(R)$ gravity that predict a not-too-heavy scalaron (not-too-small $F_{RR}$) leading to the fluctuating mode of the scalaron.

Finally, we discuss prospects for constraining the early-Universe model of $F(R)$ gravity.
We have considered only experimental bounds from tests of violation of the equivalence principle and theoretical bounds from the BBN epoch. 
However, a viable model should also satisfy other constraints from local dense regions, such as the Galactic Center, Earth, and the neutron star.
Regardless of the EDE for the Hubble tension, it is intriguing to constrain the functional form of general $F(R)$ models by combining the cosmological observations with local experiments.
This idea is not limited to $F(R)$ gravity but is valid for more general modified gravity theories.
It is worth exploring such a direction in the study of modified gravity theory.

\begin{acknowledgments}
H. C. appreciates the hospitality of the QG Lab at Nagoya University and thanks Yunwei Yu for his financial support during H. C.'s visit there.
H. C. also thanks Tomohiro Inagaki and his colleagues for their hospitality during his visit to Hiroshima University.
T. K. is supported by the National Natural Science Foundation of China (No.~12403003) and National Key R\&D Program of China (No.~2021YFA0718500).
S. N. is supported by the High-end Foreign Expert Introduction Program by the Ministry of Science and Technology of People's Republic of China (G2023158008).
T. Q. thanks the hospitality of the Institute of Theoretical Physics, Chinese Academy of Sciences, during his visit there. T.Q. is supported by the National Key Research and Development Program of China (Grant No.~2021YFC2203100), as well as Project No.~12047503 supported by the National Natural Science Foundation of China.
\end{acknowledgments}

\bibliographystyle{apsrev4-1}
\bibliography{references}

@article{Karwal:2016vyq,
    author = "Karwal, Tanvi and Kamionkowski, Marc",
    title = "{Dark energy at early times, the Hubble parameter, and the string axiverse}",
    eprint = "1608.01309",
    archivePrefix = "arXiv",
    primaryClass = "astro-ph.CO",
    doi = "10.1103/PhysRevD.94.103523",
    journal = "Phys. Rev. D",
    volume = "94",
    number = "10",
    pages = "103523",
    year = "2016"
}

@article{Poulin:2018dzj,
    author = "Poulin, Vivian and Smith, Tristan L. and Grin, Daniel and Karwal, Tanvi and Kamionkowski, Marc",
    title = "{Cosmological implications of ultralight axionlike fields}",
    eprint = "1806.10608",
    archivePrefix = "arXiv",
    primaryClass = "astro-ph.CO",
    doi = "10.1103/PhysRevD.98.083525",
    journal = "Phys. Rev. D",
    volume = "98",
    number = "8",
    pages = "083525",
    year = "2018"
}

@article{DiValentino:2021izs,
    author = "Di Valentino, Eleonora and Mena, Olga and Pan, Supriya and Visinelli, Luca and Yang, Weiqiang and Melchiorri, Alessandro and Mota, David F. and Riess, Adam G. and Silk, Joseph",
    title = "{In the realm of the Hubble tension\textemdash{}a review of solutions}",
    eprint = "2103.01183",
    archivePrefix = "arXiv",
    primaryClass = "astro-ph.CO",
    reportNumber = "IPPP/20/108",
    doi = "10.1088/1361-6382/ac086d",
    journal = "Class. Quant. Grav.",
    volume = "38",
    number = "15",
    pages = "153001",
    year = "2021"
}

@article{Schoneberg:2021qvd,
    author = {Sch\"oneberg, Nils and Franco Abell\'an, Guillermo and P\'erez S\'anchez, Andrea and Witte, Samuel J. and Poulin, Vivian and Lesgourgues, Julien},
    title = "{The H0 Olympics: A fair ranking of proposed models}",
    eprint = "2107.10291",
    archivePrefix = "arXiv",
    primaryClass = "astro-ph.CO",
    doi = "10.1016/j.physrep.2022.07.001",
    journal = "Phys. Rept.",
    volume = "984",
    pages = "1--55",
    year = "2022"
}

@article{Kamionkowski:2022pkx,
    author = "Kamionkowski, Marc and Riess, Adam G.",
    title = "{The Hubble Tension and Early Dark Energy}",
    eprint = "2211.04492",
    archivePrefix = "arXiv",
    primaryClass = "astro-ph.CO",
    doi = "10.1146/annurev-nucl-111422-024107",
    journal = "Ann. Rev. Nucl. Part. Sci.",
    volume = "73",
    pages = "153--180",
    year = "2023"
}

@article{Karwal:2021vpk,
    author = "Karwal, Tanvi and Raveri, Marco and Jain, Bhuvnesh and Khoury, Justin and Trodden, Mark",
    title = "{Chameleon early dark energy and the Hubble tension}",
    eprint = "2106.13290",
    archivePrefix = "arXiv",
    primaryClass = "astro-ph.CO",
    doi = "10.1103/PhysRevD.105.063535",
    journal = "Phys. Rev. D",
    volume = "105",
    number = "6",
    pages = "063535",
    year = "2022"
}

@article{DeFelice:2010aj,
    author = "De Felice, Antonio and Tsujikawa, Shinji",
    title = "{f(R) theories}",
    eprint = "1002.4928",
    archivePrefix = "arXiv",
    primaryClass = "gr-qc",
    doi = "10.12942/lrr-2010-3",
    journal = "Living Rev. Rel.",
    volume = "13",
    pages = "3",
    year = "2010"
}

@article{Nojiri:2017ncd,
    author = "Nojiri, S. and Odintsov, S. D. and Oikonomou, V. K.",
    title = "{Modified Gravity Theories on a Nutshell: Inflation, Bounce and Late-time Evolution}",
    eprint = "1705.11098",
    archivePrefix = "arXiv",
    primaryClass = "gr-qc",
    reportNumber = "PHYS.REPT.-692-(2017)-1-104, Phys.Rept. 692 (2017) 1-104",
    doi = "10.1016/j.physrep.2017.06.001",
    journal = "Phys. Rept.",
    volume = "692",
    pages = "1--104",
    year = "2017"
}

@article{Nojiri:2019fft,
    author = "Nojiri, Shin'ichi and Odintsov, Sergei D. and Oikonomou, V. K.",
    title = "{Unifying Inflation with Early and Late-time Dark Energy in $F(R)$ Gravity}",
    eprint = "1912.13128",
    archivePrefix = "arXiv",
    primaryClass = "gr-qc",
    doi = "10.1016/j.dark.2020.100602",
    journal = "Phys. Dark Univ.",
    volume = "29",
    pages = "100602",
    year = "2020"
}

@article{Odintsov:2020qzd,
    author = "Odintsov, Sergei D. and S\'aez-Chill\'on G\'omez, Diego and Sharov, German S.",
    title = "{Analyzing the $H_0$ tension in $F(R)$ gravity models}",
    eprint = "2011.03957",
    archivePrefix = "arXiv",
    primaryClass = "gr-qc",
    doi = "10.1016/j.nuclphysb.2021.115377",
    journal = "Nucl. Phys. B",
    volume = "966",
    pages = "115377",
    year = "2021"
}

@article{Odintsov:2023cli,
    author = "Odintsov, Sergei D. and Oikonomou, V. K. and Sharov, German S.",
    title = "{Early dark energy with power-law F(R) gravity}",
    eprint = "2305.17513",
    archivePrefix = "arXiv",
    primaryClass = "gr-qc",
    doi = "10.1016/j.physletb.2023.137988",
    journal = "Phys. Lett. B",
    volume = "843",
    pages = "137988",
    year = "2023"
}

@article{Nojiri:2022ski,
    author = "Nojiri, S. and Odintsov, S. D. and Oikonomou, V. K.",
    title = "{Integral F(R) gravity and saddle point condition as a remedy for the H0-tension}",
    eprint = "2205.11681",
    archivePrefix = "arXiv",
    primaryClass = "gr-qc",
    doi = "10.1016/j.nuclphysb.2022.115850",
    journal = "Nucl. Phys. B",
    volume = "980",
    pages = "115850",
    year = "2022"
}

@article{Schiavone:2022wvq,
    author = "Schiavone, Tiziano and Montani, Giovanni and Bombacigno, Flavio",
    title = "{f(R) gravity in the Jordan frame as a paradigm for the Hubble tension}",
    eprint = "2211.16737",
    archivePrefix = "arXiv",
    primaryClass = "gr-qc",
    doi = "10.1093/mnrasl/slad041",
    journal = "Mon. Not. Roy. Astron. Soc.",
    volume = "522",
    number = "1",
    pages = "L72--L77",
    year = "2023"
}

@article{Dainotti:2021pqg,
    author = "Dainotti, Maria Giovanna and De Simone, Biagio and Schiavone, Tiziano and Montani, Giovanni and Rinaldi, Enrico and Lambiase, Gaetano",
    title = "{On the Hubble constant tension in the SNe Ia Pantheon sample}",
    eprint = "2103.02117",
    archivePrefix = "arXiv",
    primaryClass = "astro-ph.CO",
    reportNumber = "AAS28095R4, RIKEN-iTHEMS-Report-21",
    doi = "10.3847/1538-4357/abeb73",
    journal = "Astrophys. J.",
    volume = "912",
    number = "2",
    pages = "150",
    year = "2021"
}

@article{Dainotti:2022bzg,
    author = "Dainotti, Maria Giovanna and De Simone, Biagio and Schiavone, Tiziano and Montani, Giovanni and Rinaldi, Enrico and Lambiase, Gaetano and Bogdan, Malgorzata and Ugale, Sahil",
    title = "{On the Evolution of the Hubble Constant with the SNe Ia Pantheon Sample and Baryon Acoustic Oscillations: A Feasibility Study for GRB-Cosmology in 2030}",
    eprint = "2201.09848",
    archivePrefix = "arXiv",
    primaryClass = "astro-ph.CO",
    reportNumber = "RIKEN-ITHEMS-Report-22",
    doi = "10.3390/galaxies10010024",
    journal = "Galaxies",
    volume = "10",
    number = "1",
    pages = "24",
    year = "2022"
}

@article{Brax:2004qh,
    author = "Brax, Philippe and van de Bruck, Carsten and Davis, Anne-Christine and Khoury, Justin and Weltman, Amanda",
    title = "{Detecting dark energy in orbit: The cosmological chameleon}",
    eprint = "astro-ph/0408415",
    archivePrefix = "arXiv",
    doi = "10.1103/PhysRevD.70.123518",
    journal = "Phys. Rev. D",
    volume = "70",
    pages = "123518",
    year = "2004"
}

@article{Chen:2022zkc,
    author = "Chen, Hua and Katsuragawa, Taishi and Matsuzaki, Shinya",
    title = "{Towards a unified interpretation of the early Universe in R $^{2}$-corrected dark energy model of F(R) gravity*}",
    eprint = "2206.02130",
    archivePrefix = "arXiv",
    primaryClass = "gr-qc",
    doi = "10.1088/1674-1137/ac7d46",
    journal = "Chin. Phys. C",
    volume = "46",
    number = "10",
    pages = "105106",
    year = "2022"
}

@article{Starobinsky:1980te,
    author = "Starobinsky, Alexei A.",
    editor = "Khalatnikov, I. M. and Mineev, V. P.",
    title = "{A New Type of Isotropic Cosmological Models Without Singularity}",
    doi = "10.1016/0370-2693(80)90670-X",
    journal = "Phys. Lett. B",
    volume = "91",
    pages = "99--102",
    year = "1980"
}

@article{Scali:2024ftw,
    author = "Scali, Federico and Piattella, Oliver Fabio",
    title = "{Asymptotically Schwarzschild solutions in f(R) extension of general relativity}",
    eprint = "2406.04417",
    archivePrefix = "arXiv",
    primaryClass = "gr-qc",
    doi = "10.1103/PhysRevD.110.064042",
    journal = "Phys. Rev. D",
    volume = "110",
    number = "6",
    pages = "064042",
    year = "2024"
}

@article{Lee:2020zjt,
    author = "Lee, J. G. and Adelberger, E. G. and Cook, T. S. and Fleischer, S. M. and Heckel, B. R.",
    title = "{New Test of the Gravitational $1/r^2$ Law at Separations down to 52 $\mu$m}",
    eprint = "2002.11761",
    archivePrefix = "arXiv",
    primaryClass = "hep-ex",
    doi = "10.1103/PhysRevLett.124.101101",
    journal = "Phys. Rev. Lett.",
    volume = "124",
    number = "10",
    pages = "101101",
    year = "2020"
}

@article{Capozziello:2007eu,
    author = "Capozziello, Salvatore and Tsujikawa, Shinji",
    title = "{Solar system and equivalence principle constraints on f(R) gravity by chameleon approach}",
    eprint = "0712.2268",
    archivePrefix = "arXiv",
    primaryClass = "gr-qc",
    doi = "10.1103/PhysRevD.77.107501",
    journal = "Phys. Rev. D",
    volume = "77",
    pages = "107501",
    year = "2008"
}

@article{Lin:2019qug,
    author = "Lin, Meng-Xiang and Benevento, Giampaolo and Hu, Wayne and Raveri, Marco",
    title = "{Acoustic Dark Energy: Potential Conversion of the Hubble Tension}",
    eprint = "1905.12618",
    archivePrefix = "arXiv",
    primaryClass = "astro-ph.CO",
    doi = "10.1103/PhysRevD.100.063542",
    journal = "Phys. Rev. D",
    volume = "100",
    number = "6",
    pages = "063542",
    year = "2019"
}

@article{Niedermann:2019olb,
    author = "Niedermann, Florian and Sloth, Martin S.",
    title = "{New early dark energy}",
    eprint = "1910.10739",
    archivePrefix = "arXiv",
    primaryClass = "astro-ph.CO",
    doi = "10.1103/PhysRevD.103.L041303",
    journal = "Phys. Rev. D",
    volume = "103",
    number = "4",
    pages = "L041303",
    year = "2021"
}

@article{Braglia:2020bym,
    author = "Braglia, Matteo and Emond, William T. and Finelli, Fabio and Gumrukcuoglu, A. Emir and Koyama, Kazuya",
    title = "{Unified framework for early dark energy from $\alpha$-attractors}",
    eprint = "2005.14053",
    archivePrefix = "arXiv",
    primaryClass = "astro-ph.CO",
    doi = "10.1103/PhysRevD.102.083513",
    journal = "Phys. Rev. D",
    volume = "102",
    number = "8",
    pages = "083513",
    year = "2020"
}

@article{Jiang:2021bab,
    author = "Jiang, Jun-Qian and Piao, Yun-Song",
    title = "{Testing AdS early dark energy with Planck, SPTpol, and LSS data}",
    eprint = "2107.07128",
    archivePrefix = "arXiv",
    primaryClass = "astro-ph.CO",
    doi = "10.1103/PhysRevD.104.103524",
    journal = "Phys. Rev. D",
    volume = "104",
    number = "10",
    pages = "103524",
    year = "2021"
}

@article{Sabla:2022xzj,
    author = "Sabla, Vivian I. and Caldwell, Robert R.",
    title = "{Microphysics of early dark energy}",
    eprint = "2202.08291",
    archivePrefix = "arXiv",
    primaryClass = "astro-ph.CO",
    doi = "10.1103/PhysRevD.106.063526",
    journal = "Phys. Rev. D",
    volume = "106",
    number = "6",
    pages = "063526",
    year = "2022"
}

@article{McDonough:2021pdg,
    author = "McDonough, Evan and Lin, Meng-Xiang and Hill, J. Colin and Hu, Wayne and Zhou, Shengjia",
    title = "{Early dark sector, the Hubble tension, and the swampland}",
    eprint = "2112.09128",
    archivePrefix = "arXiv",
    primaryClass = "astro-ph.CO",
    doi = "10.1103/PhysRevD.106.043525",
    journal = "Phys. Rev. D",
    volume = "106",
    number = "4",
    pages = "043525",
    year = "2022"
}

@article{Lin:2022phm,
    author = "Lin, Meng-Xiang and McDonough, Evan and Hill, J. Colin and Hu, Wayne",
    title = "{Dark matter trigger for early dark energy coincidence}",
    eprint = "2212.08098",
    archivePrefix = "arXiv",
    primaryClass = "astro-ph.CO",
    doi = "10.1103/PhysRevD.107.103523",
    journal = "Phys. Rev. D",
    volume = "107",
    number = "10",
    pages = "103523",
    year = "2023"
}

@article{Lima:2016npg,
    author = "Lima, Nelson A. and Smer-Barreto, Vanessa and Lombriser, Lucas",
    title = "{Constraints on decaying early modified gravity from cosmological observations}",
    eprint = "1603.05239",
    archivePrefix = "arXiv",
    primaryClass = "astro-ph.CO",
    doi = "10.1103/PhysRevD.94.083507",
    journal = "Phys. Rev. D",
    volume = "94",
    number = "8",
    pages = "083507",
    year = "2016"
}

@article{Benevento:2022cql,
    author = "Benevento, Giampaolo and Kable, Joshua A. and Addison, Graeme E. and Bennett, Charles L.",
    title = "{An Exploration of an Early Gravity Transition in Light of Cosmological Tensions}",
    eprint = "2202.09356",
    archivePrefix = "arXiv",
    primaryClass = "astro-ph.CO",
    doi = "10.3847/1538-4357/ac80fd",
    journal = "Astrophys. J.",
    volume = "935",
    number = "2",
    pages = "156",
    year = "2022"
}

@article{Lin:2018nxe,
    author = "Lin, Meng-Xiang and Raveri, Marco and Hu, Wayne",
    title = "{Phenomenology of Modified Gravity at Recombination}",
    eprint = "1810.02333",
    archivePrefix = "arXiv",
    primaryClass = "astro-ph.CO",
    doi = "10.1103/PhysRevD.99.043514",
    journal = "Phys. Rev. D",
    volume = "99",
    number = "4",
    pages = "043514",
    year = "2019"
}

@article{SolaPeracaula:2019zsl,
    author = "Sol\`a Peracaula, Joan and Gomez-Valent, Adria and de Cruz P\'erez, Javier and Moreno-Pulido, Cristian",
    title = "{Brans\textendash{}Dicke Gravity with a Cosmological Constant Smoothes Out $\Lambda$CDM Tensions}",
    eprint = "1909.02554",
    archivePrefix = "arXiv",
    primaryClass = "astro-ph.CO",
    doi = "10.3847/2041-8213/ab53e9",
    journal = "Astrophys. J. Lett.",
    volume = "886",
    number = "1",
    pages = "L6",
    year = "2019"
}

@article{Ballardini:2020iws,
    author = "Ballardini, Mario and Braglia, Matteo and Finelli, Fabio and Paoletti, Daniela and Starobinsky, Alexei A. and Umilt\`a, Caterina",
    title = "{Scalar-tensor theories of gravity, neutrino physics, and the $H_0$ tension}",
    eprint = "2004.14349",
    archivePrefix = "arXiv",
    primaryClass = "astro-ph.CO",
    doi = "10.1088/1475-7516/2020/10/044",
    journal = "JCAP",
    volume = "10",
    pages = "044",
    year = "2020"
}

@article{Ballardini:2021evv,
    author = "Ballardini, Mario and Finelli, Fabio and Sapone, Domenico",
    title = "{Cosmological constraints on the gravitational constant}",
    eprint = "2111.09168",
    archivePrefix = "arXiv",
    primaryClass = "astro-ph.CO",
    doi = "10.1088/1475-7516/2022/06/004",
    journal = "JCAP",
    volume = "06",
    number = "06",
    pages = "004",
    year = "2022"
}

@article{Ballesteros:2020sik,
    author = "Ballesteros, Guillermo and Notari, Alessio and Rompineve, Fabrizio",
    title = "{The $H_0$ tension: $\Delta G_N$ vs. $\Delta N_{\rm eff}$}",
    eprint = "2004.05049",
    archivePrefix = "arXiv",
    primaryClass = "astro-ph.CO",
    doi = "10.1088/1475-7516/2020/11/024",
    journal = "JCAP",
    volume = "11",
    pages = "024",
    year = "2020"
}

@article{Zumalacarregui:2020cjh,
    author = "Zumalacarregui, Miguel",
    title = "{Gravity in the Era of Equality: Towards solutions to the Hubble problem without fine-tuned initial conditions}",
    eprint = "2003.06396",
    archivePrefix = "arXiv",
    primaryClass = "astro-ph.CO",
    doi = "10.1103/PhysRevD.102.023523",
    journal = "Phys. Rev. D",
    volume = "102",
    number = "2",
    pages = "023523",
    year = "2020"
}

@article{Braglia:2020auw,
    author = "Braglia, Matteo and Ballardini, Mario and Finelli, Fabio and Koyama, Kazuya",
    title = "{Early modified gravity in light of the $H_0$ tension and LSS data}",
    eprint = "2011.12934",
    archivePrefix = "arXiv",
    primaryClass = "astro-ph.CO",
    doi = "10.1103/PhysRevD.103.043528",
    journal = "Phys. Rev. D",
    volume = "103",
    number = "4",
    pages = "043528",
    year = "2021"
}

@article{Benevento:2020fev,
    author = "Benevento, Giampaolo and Hu, Wayne and Raveri, Marco",
    title = "{Can Late Dark Energy Transitions Raise the Hubble constant?}",
    eprint = "2002.11707",
    archivePrefix = "arXiv",
    primaryClass = "astro-ph.CO",
    doi = "10.1103/PhysRevD.101.103517",
    journal = "Phys. Rev. D",
    volume = "101",
    number = "10",
    pages = "103517",
    year = "2020"
}

@article{Kable:2023bsg,
    author = "Kable, Joshua A. and Benevento, Giampaolo and Addison, Graeme E. and Bennett, Charles L.",
    title = "{Cosmological Tensions and the Transitional Planck Mass Model}",
    eprint = "2307.12174",
    archivePrefix = "arXiv",
    primaryClass = "astro-ph.CO",
    doi = "10.3847/1538-4357/acfed0",
    journal = "Astrophys. J.",
    volume = "959",
    number = "2",
    pages = "143",
    year = "2023"
}

@article{Erickcek:2013oma,
    author = "Erickcek, Adrienne L. and Barnaby, Neil and Burrage, Clare and Huang, Zhiqi",
    title = "{Catastrophic Consequences of Kicking the Chameleon}",
    eprint = "1304.0009",
    archivePrefix = "arXiv",
    primaryClass = "astro-ph.CO",
    doi = "10.1103/PhysRevLett.110.171101",
    journal = "Phys. Rev. Lett.",
    volume = "110",
    pages = "171101",
    year = "2013"
}

@article{Erickcek:2013dea,
    author = "Erickcek, Adrienne L. and Barnaby, Neil and Burrage, Clare and Huang, Zhiqi",
    title = "{Chameleons in the Early Universe: Kicks, Rebounds, and Particle Production}",
    eprint = "1310.5149",
    archivePrefix = "arXiv",
    primaryClass = "astro-ph.CO",
    doi = "10.1103/PhysRevD.89.084074",
    journal = "Phys. Rev. D",
    volume = "89",
    number = "8",
    pages = "084074",
    year = "2014"
}

@article{Katsuragawa:2017wge,
    author = "Katsuragawa, Taishi and Matsuzaki, Shinya",
    title = "{Cosmic History of Chameleonic Dark Matter in $F(R)$ Gravity}",
    eprint = "1708.08702",
    archivePrefix = "arXiv",
    primaryClass = "gr-qc",
    doi = "10.1103/PhysRevD.97.064037",
    journal = "Phys. Rev. D",
    volume = "97",
    number = "6",
    pages = "064037",
    year = "2018",
    note = "[Erratum: Phys.Rev.D 97, 129902 (2018)]"
}

@article{Katsuragawa:2018wbe,
    author = "Katsuragawa, Taishi and Matsuzaki, Shinya and Senaha, Eibun",
    title = "{$F(R)$ gravity in the early Universe: Electroweak phase transition and chameleon mechanism}",
    eprint = "1812.00640",
    archivePrefix = "arXiv",
    primaryClass = "gr-qc",
    doi = "10.1088/1674-1137/43/10/105101",
    journal = "Chin. Phys. C",
    volume = "43",
    number = "10",
    pages = "105101",
    year = "2019"
}

@article{Chen:2019kcu,
    author = "Chen, Hua and Katsuragawa, Taishi and Matsuzaki, Shinya and Qiu, Taotao",
    title = "{Big Bang Nucleosynthesis Hunts Chameleon Dark Matter}",
    eprint = "1908.04146",
    archivePrefix = "arXiv",
    primaryClass = "hep-ph",
    doi = "10.1007/JHEP02(2020)155",
    journal = "JHEP",
    volume = "02",
    pages = "155",
    year = "2020"
}

@article{Shtanov:2021uif,
    author = "Shtanov, Yuri",
    title = "{Light scalaron as dark matter}",
    eprint = "2105.02662",
    archivePrefix = "arXiv",
    primaryClass = "hep-ph",
    doi = "10.1016/j.physletb.2021.136469",
    journal = "Phys. Lett. B",
    volume = "820",
    pages = "136469",
    year = "2021"
}

@article{Shtanov:2022xew,
    author = "Shtanov, Yuri",
    title = "{Initial conditions for the scalaron dark matter}",
    eprint = "2207.00267",
    archivePrefix = "arXiv",
    primaryClass = "astro-ph.CO",
    doi = "10.1088/1475-7516/2022/10/079",
    journal = "JCAP",
    volume = "10",
    pages = "079",
    year = "2022"
}

@article{Shtanov:2024nmf,
    author = "Shtanov, Yuri",
    title = "{Scalaron dark matter and the thermal history of the universe}",
    eprint = "2409.05027",
    archivePrefix = "arXiv",
    primaryClass = "hep-ph",
    month = "9",
    year = "2024"
}

@article{Numajiri:2021nsc,
    author = "Numajiri, Kota and Katsuragawa, Taishi and Nojiri, Shin'ichi",
    title = "{Compact star in general F(R) gravity: Inevitable degeneracy problem and non-integer power correction}",
    eprint = "2111.02660",
    archivePrefix = "arXiv",
    primaryClass = "gr-qc",
    doi = "10.1016/j.physletb.2022.136929",
    journal = "Phys. Lett. B",
    volume = "826",
    pages = "136929",
    year = "2022"
}

@article{Codello:2014sua,
    author = "Codello, Alessandro and Joergensen, Jakob and Sannino, Francesco and Svendsen, Ole",
    title = "{Marginally Deformed Starobinsky Gravity}",
    eprint = "1404.3558",
    archivePrefix = "arXiv",
    primaryClass = "hep-ph",
    reportNumber = "CP3-ORIGINS-2014-13, DIAS-2014-13",
    doi = "10.1007/JHEP02(2015)050",
    journal = "JHEP",
    volume = "02",
    pages = "050",
    year = "2015"
}

@article{Hu:2007nk,
      author         = "Hu, Wayne and Sawicki, Ignacy",
      title          = "{Models of f(R) Cosmic Acceleration that Evade
                        Solar-System Tests}",
      journal        = "Phys. Rev.",
      volume         = "D76",
      year           = "2007",
      pages          = "064004",
      doi            = "10.1103/PhysRevD.76.064004",
      eprint         = "0705.1158",
      archivePrefix  = "arXiv",
      primaryClass   = "astro-ph",
      SLACcitation   = "%%CITATION = ARXIV:0705.1158;%%"
}

@article{Starobinsky:2007hu,
      author         = "Starobinsky, Alexei A.",
      title          = "{Disappearing cosmological constant in f(R) gravity}",
      journal        = "JETP Lett.",
      volume         = "86",
      year           = "2007",
      pages          = "157-163",
      doi            = "10.1134/S0021364007150027",
      eprint         = "0706.2041",
      archivePrefix  = "arXiv",
      primaryClass   = "astro-ph",
      SLACcitation   = "%%CITATION = ARXIV:0706.2041;%%"
}

@article{Appleby:2007vb,
      author         = "Appleby, Stephen A. and Battye, Richard A.",
      title          = "{Do consistent $F(R)$ models mimic General Relativity
                        plus $\Lambda$?}",
      journal        = "Phys. Lett.",
      volume         = "B654",
      year           = "2007",
      pages          = "7-12",
      doi            = "10.1016/j.physletb.2007.08.037",
      eprint         = "0705.3199",
      archivePrefix  = "arXiv",
      primaryClass   = "astro-ph",
      SLACcitation   = "%%CITATION = ARXIV:0705.3199;%%"
}

@article{Tsujikawa:2007xu,
      author         = "Tsujikawa, Shinji",
      title          = "{Observational signatures of $f(R)$ dark energy models
                        that satisfy cosmological and local gravity constraints}",
      journal        = "Phys. Rev.",
      volume         = "D77",
      year           = "2008",
      pages          = "023507",
      doi            = "10.1103/PhysRevD.77.023507",
      eprint         = "0709.1391",
      archivePrefix  = "arXiv",
      primaryClass   = "astro-ph",
      SLACcitation   = "%%CITATION = ARXIV:0709.1391;%%"
}

@article{Elizalde:2010ts,
    author = "Elizalde, E. and Nojiri, S. and Odintsov, S. D. and Sebastiani, L. and Zerbini, S.",
    title = "{Non-singular exponential gravity: a simple theory for early- and late-time accelerated expansion}",
    eprint = "1012.2280",
    archivePrefix = "arXiv",
    primaryClass = "hep-th",
    doi = "10.1103/PhysRevD.83.086006",
    journal = "Phys. Rev. D",
    volume = "83",
    pages = "086006",
    year = "2011"
}

\end{document}